\documentclass{article}
\usepackage{graphicx} 
\usepackage{amsmath,amssymb}
\usepackage{amsfonts}
\usepackage{amsthm}
\usepackage{color}
\usepackage{url}
\usepackage{geometry}
\DeclareMathOperator{\diag}{\mathrm{diag}}

\newcommand{\Lap}{\bigtriangleup}
\newcommand{\dd}{\mathrm{d}}
\newcommand{\e}{\mathrm{e}}

\title{Oscillations in neuronal activity: a neuron-centered spatiotemporal model of the Unfolded Protein Response in prion diseases}
\author{Elliot M. Miller \thanks{College of Arts and Sciences, Culverhouse College of Business, The University of Alabama, Tuscaloosa, AL, USA} \and Tat Chung D. Chan \thanks{Department of Mathematics, University of California Berkeley, Berkeley, CA, USA} \and Carlos Montes-Matamoros \thanks{School of Mathematical and Statistical Sciences, The University of Texas Rio Grande Valley, Edinburg, TX, USA} \and Omar Sharif \footnotemark[3] \and  Laurent Pujo-Menjouet \thanks{Universite Claude Bernard Lyon 1, CNRS, Ecole Centrale de Lyon, INSA Lyon, Université Jean Monnet, ICJ UMR5208, Inria, 69622 Villeurbanne, France} \and Michael R. Lindstrom \footnotemark[3]}

\date{\today}

\begin{document}

\maketitle

\begin{abstract}
    Many neurodegenerative diseases (NDs) are characterized by the slow spatial spread of toxic protein species in the brain. The toxic proteins can induce neuronal stress, triggering the Unfolded Protein Response (UPR), which slows or stops protein translation and can indirectly reduce the toxic load. However, the UPR may also trigger processes leading to apoptotic cell death and the UPR is implicated in the progression of several NDs. In this paper, we develop a novel mathematical model to describe the spatiotemporal dynamics of the UPR mechanism for    prion diseases. Our model is centered  around a single neuron, with representative proteins P (healthy) and S (toxic) interacting with heterodimer dynamics (S interacts with P to form two S's). The model takes the form of a coupled system of nonlinear reaction-diffusion equations with a delayed, nonlinear flux for P (delay from the UPR). Through the delay, we find parameter regimes that exhibit oscillations in the P- and S-protein levels. We find that oscillations are more pronounced when the S-clearance rate and S-diffusivity are small in comparison to the P-clearance rate and P-diffusivity, respectively. The oscillations become more pronounced as delays in initiating the UPR increase. We also consider quasi-realistic clinical parameters to understand how possible drug therapies can alter the course of a prion disease. We find that decreasing the production of P, decreasing the recruitment rate, increasing the diffusivity of S, increasing the UPR S-threshold, and increasing the S clearance rate appear to be the most powerful modifications to reduce the mean UPR intensity and potentially moderate the disease progression. 
\end{abstract}


\section{Introduction}

Neurodegenerative diseases (NDs) are devastating conditions affecting the brain and central nervous system. Two of the most common NDs are Alzheimer's Disease \cite{goedert2006century}, with the gradual loss of memory and capacity to function independently, and Parkinson's Disease \cite{davie2008review}, with loss of muscle coordination and cognitive impairment. While various hypotheses exist as to the origins and etiologies of NDs, medicine is not yet able to cure most of them \cite{akhtar2021neurodegenerative}. 

Despite their vastly different presentations and affected brain regions, many NDs share a common set of features. The main commonality is the presence of toxic proteins, that are thought to impair neuronal function or cause cell death, spreading through the brain \cite{kiaei2013new}. During this spreading, there is a recruitment process where normal proteins can become toxic through interacting with other toxic proteins in prion diseases \cite{newby2013blessings} or can become toxic through forming various oligomers \cite{ono2009structure}. In the case of AD, the toxic proteins are thought to be Amyloid-beta oligomers \cite{ono2009structure,tolar2020path}; in Parkinson's, Alpha-synuclein\cite{stefanis2012alpha}; in prion diseases, Scrapie proteins \cite{prusiner1989creutzfeldt}; and so on.

When subjected to the stress of a toxin, neurons are known to exhibit the Unfolded Protein Response (UPR) \cite{hetz2020mechanisms} and may temporarily shut down or limit their cellular processes, reducing their production of healthy, endogenous proteins \cite{halliday2014targeting}. This limiting thereby cuts down on the supply chain that could yield more toxic proteins, and allows natural clearance mechanisms of the toxins to take place. When the stress is lowered, the neuron can commence its normal cellular functions again. The activation of the UPR may also cause damage and apoptosis to the neurons. When activated, the UPR results in increasing production of the endoplasmic reticulum (ER) kinases, PERK and Ire1, which reduce protein translation. But, these proteins also lead to apoptosis through a cascade of processes \cite{fricker2018neuronal}. Post-mortem autopsies of ND patients suggest the UPR has been activated \cite{halliday2014targeting}. It is also suggested that weakened UPR mechanisms may be implicated in the progression of NDs \cite{apodaca2006cellular}.

Given the innate complexities of biological systems, it is extremely difficult to study how individual mechanisms affect disease etiologies. This is where mathematical modelling can be extremely useful, giving researchers a mechanism to put together a set of assumptions and to observe the outcomes of model systems subject to those assumptions. Mathematical models have been used to study a host of different disorders and mechanisms, including AD \cite{hao2016mathematical,puri2010mathematical,lindstrom2021reaction}, PD\cite{desplats2009inclusion,bakshi2019mathematical,pandya2019predictive}, CJD \cite{greer2006mathematical,salman2018mathematical}, the UPR \cite{adimy2022neuron,adimy2023multigroup,trusina2010unfolded}, and many others. The models considered cover a range of scales from studies of a single neuron or two \cite{adimy2022neuron}, to the spatial spreading of toxic proteins in the brain through modelling the connectome and how it evolves \cite{goriely2020neuronal}.

In recent work, a simple compartment model, comprised of a nonlinear system of Delay Differential Equations (DDEs), was introduced to model the UPR in prion diseases \cite{adimy2022neuron}. This yielded intriguing results whereby the presence of delay was able to induce oscillations in the levels of toxic proteins, and these oscillations could be turned on/off in different parameter regimes. Moreover, the parameters could be tuned to drive the toxic protein concentration to zero. 

In this paper, our objective is to extend the prior model of \cite{adimy2022neuron} to a delayed spatiotemporal model. We develop a nonlinear system of Reaction Diffusion Equations, with a nonlinear flux term exhibiting a delay. Our model focuses on prion diseases using heterodimer dynamics \cite{garzon2021dynamics}, centered around a single neuron. Cellular prion protein, PrP$^\text{C}$ is produced by neurons; through its interaction with misfolded, toxic scrapie prion proteins, PrP$^\text{Sc}$, PrP$^\text{C}$ misfolds into PrP$^\text{Sc}$ \cite{atkinson2016prion, selkoe2003folding}. We numerically investigate parameters that yield oscillations and which parameter modifications can reduce select measures of disease severity.

Our paper is organized as follows: we present our method of study, a mathematical model, in Section \ref{sec:mm}; an exploration of the model and its general behavior is covered in Section \ref{sec:results}; we focus on the biological parameters and treatment implications for prion diseases in Section \ref{sec:bio}; and our work is concluded in Section \ref{sec:conclusion}. The Appendices \ref{app:num} and \ref{app:spec} provide additional mathematical details not located in the main text.

\section{Materials and Methods}

\label{sec:mm}

We focus on gaining an understanding into the UPR mechanism and its relation to prion diseases at the neuron scale. In isolation, different mechanisms have been studied experimentally; however, at present, an in-depth knowledge of {\it in vivo} parameters of biological significance is lacking. For this reason, we combine multiple well-established biological processes into a {\it mathematical model}, to simulate a system and glean understanding into its dynamics. The hope is that enough of the important biology is present in the model for it to provide clinically relevant insights, even if only in approximation. Our experiments then take the form of numerical simulations, where each parameter can be carefully controlled.

A reader less focused on the mathematics, with more interest in the biology, should consider reading Section \ref{sec:overview} and then reading Sections \ref{sec:results} and \ref{sec:bio}. It would also be useful to refer to Table \ref{tab:allparams} to see the different parameters studied in our model. 

\subsection{Model}

This section concerns developing a mathematical model.

\subsubsection{Overview}

\label{sec:overview}

Our objectives with this model are to uncover the spatiotemporal dynamics of the Unfolded Protein Response. For simplicity, we consider two representative protein species, P (healthy proteins, different forms of the cellular prion protein, PrP$^\text{C}$) and S (misfolded proteins, different forms of the toxic scrapie protein, PrP$^\text{Sc}$). The S-proteins can recruit P-proteins to become S-proteins. We do not model higher-order structures --- dimers, higher-order oligomers, nuclei, and fibrils are not present. We argue that the capacity for S to recruit P is at least representative of the more detailed biochemistry whereby misfolded proteins can form oligomers and nuclei, which can fragment to generate further recruitment. This is often referred to as a heterodimer model.

Our model centers around one neuron, with dynamics taking place in the interstitial fluid surrounding it. The P-proteins are produced within the neuron and released through the membrane into the intercellular space. We note cellular prion protein and scrapie protein can be found as both membrane-bound and extracellular forms \cite{shafiq2022prion,rangel2013non}, but we focus upon the latter here to build a simple model. Through a buildup of S near the membrane, the P-production is lowered (due to the UPR) after some delay representing the processing time needed to stop/slow translation. 

Figure \ref{fig:math_model} depicts the relevant mechanisms. Our simulations center around the geometries depicted in Figure \ref{fig:symmetry}.

\subsubsection{Derivation}

We denote $P$ the concentration of P-proteins and $S$ the concentration of S-proteins. We use $x$ for position and $t$ for time.

From a chemical reaction perspective, we assume that
\begin{itemize}
    \item P and S combine at a rate $c \geq 0$ to form two separate S proteins; 
    \item P is cleared at a rate $a \geq 0$; 
    \item S is cleared at a rate $b \geq 0$, with $b \leq a$; 
    \item spontaneously, P can misfold to become S at a rate $f \geq 0$; and 
    \item P and S have diffusivities within the interneuronal space of $D_P$ and $D_S$, respectively, with $0 \leq D_S \leq D_P$.
    \end{itemize}
    The fact that $D_S \leq D_P$ stems from the fact that we expect, in general, a single S-protein is at least as massive as a single P-protein (recall we are not directly modelling oligomers but S represents misfolded proteins of all sizes). We also anticipate that $b \leq a$ as larger proteins likely take longer to break down. Given the relative rarity of prion diseases, we expect that $f \ll a$. 

To account for the UPR mechanism, we prescribe that the rate that P-proteins are produced and released decreases as the S concentration at the cell membrane increases. More precisely, the magnitude of the flux, $J,$ of P from the neuron on the neuron boundary at time $t$ is given by the proportionality relation \begin{equation} J \propto (1+(\langle S(\cdot,t-t_d) \rangle/S_c)^m)^{-1},\end{equation} with the constant of proportionality $A>0$ being the maximum possible P-flux and where $S_c>0$, $m>0$, and $t_d \geq 0$ are prescribed constants. The $\langle \cdot \rangle$ denotes the mean value over the membrane. The value $S_c$ is more or less a sensitivity of the neurons to S: when $\langle S \rangle >S_c$, the flux may decrease very rapidly and when $\langle S \rangle <S_c$, the flux stays near its maximum. The value $m$ controls how rapidly the neuron switches from maximum P-flux to near zero P-flux. The delay $t_d$ models the fact that during stress, the P-production cannot shut down immediately; a cascade of signals needs to be transmitted to decrease P-production, which is modelled by $t_d$. The same principle applies to increasing P production when S is cleared. 

We consider the system in $d$-dimensional space. While $d=3$ is most natural, certain geometric arrangements of neurons could be modelled as being $1$ or $2$ dimensional. We let the neuron occupy a bounded, closed, and connected region $\Theta \subset \mathbb{R}^d$ containing the origin. In $d=1$ we assume, without loss of generality, that $\Theta = [-R,R]$ for some length $R>0$. For $d=2, 3$, we further assume a smooth boundary and that a characteristic length scale for $\Theta$ is $R>0$. The computational domain is the region $\Omega := \mathbb{R}^d \setminus \Theta$ and the cell membrane is given by $\partial \Theta = \partial \Omega$. With $d=1$, we only model the region $(R,\infty)$ so that $\langle S(\cdot,t-t_d) \rangle = S(R, t-t_d)$. For $d=2,3$, $\langle S(\cdot,t-t_d) \rangle = \frac{1}{|\partial \Omega|} \int_{\partial \Omega} S(x, t-t_d) \dd x.$

For boundary conditions, we assume that $P$ and $S$ tend to $0$ as $|x| \to \infty$ (exponential decay is common in diffusion problems) and that there is no flux of S on the boundary $\partial \Omega$. The flux of P on $\partial \Omega$ is based on the UPR assumptions outlined above.

\begin{figure}
\centering
    \includegraphics[width=0.4\linewidth]{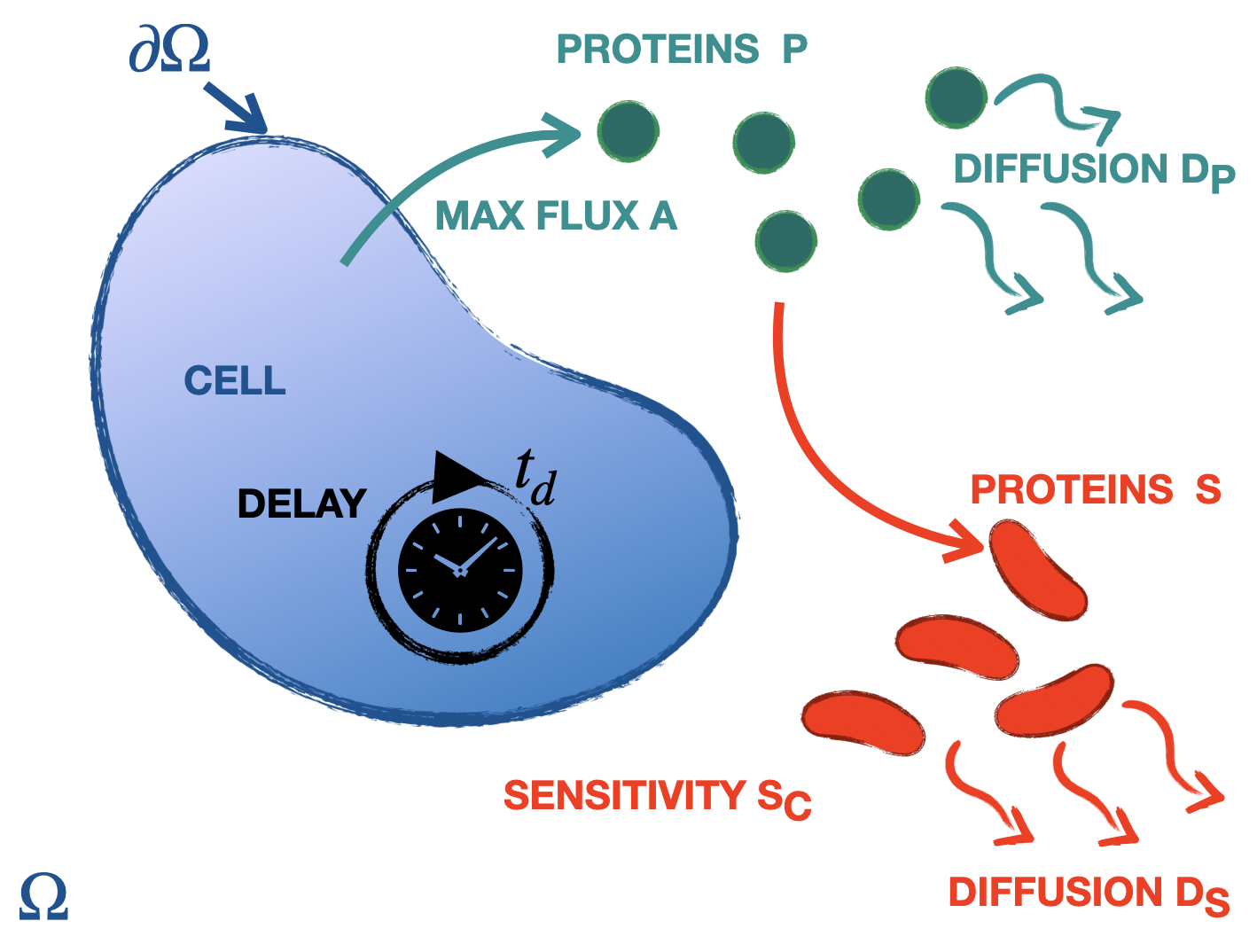} \includegraphics[width=0.4\linewidth]{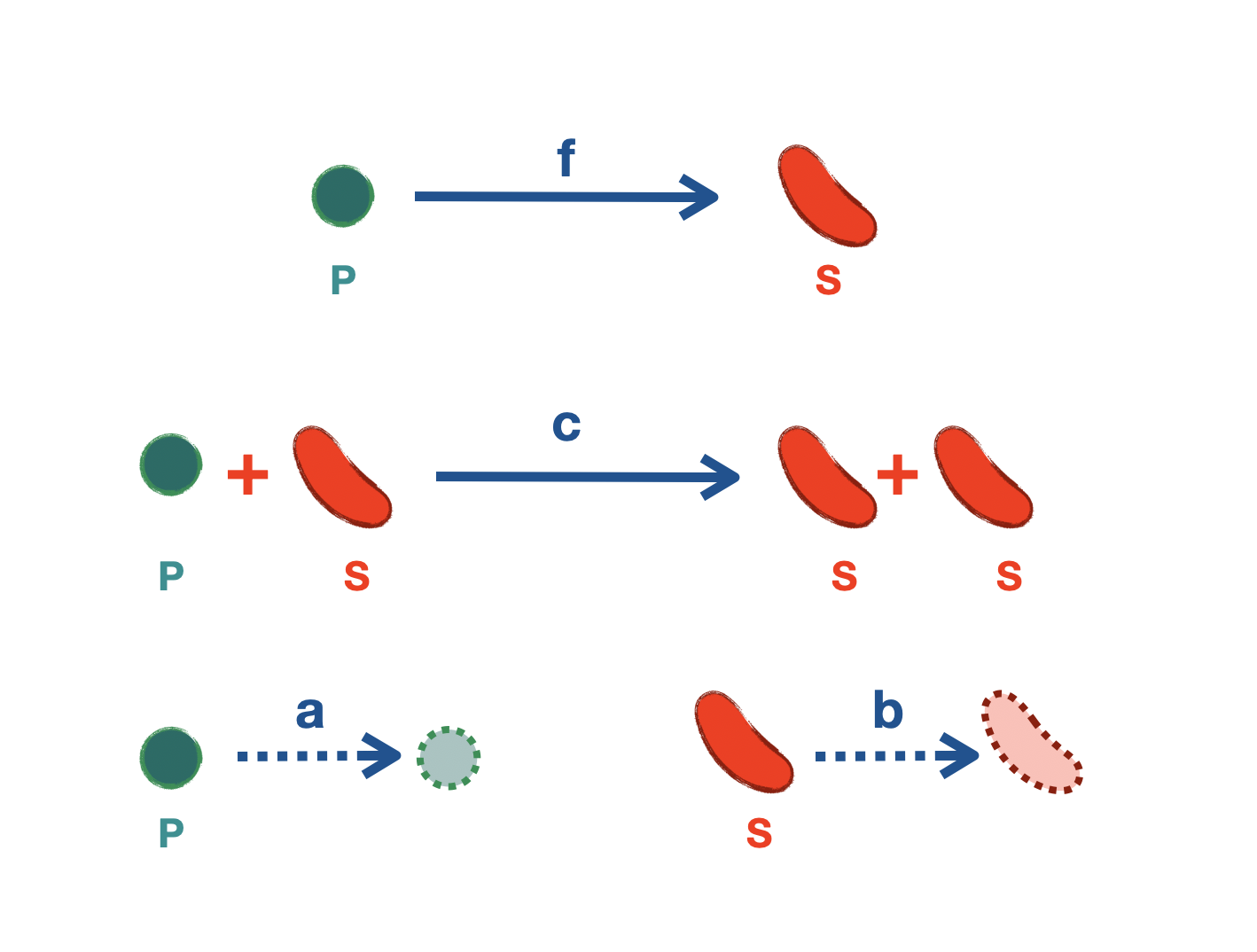} 
    \caption{Representative sketch of model. \textbf{ Left:} a cell produces protein P that is released into the intercellular space and diffuses with a rate $D_P$. P may be recruited to become S that diffuses with a rate $D_S$. The neuron is sensitive to S at a concentration $S_c$ and will reduce its maximum flux of P from $A$ in the presence of S. \textbf{Right} Spontaneous misfolding from P- to S-proteins occurs at a rate $f$, recruitment by interaction with an S-protein happens with a rate $c,$ and degradation rates for P- and S-proteins are respectively $a$ and $b$.} 
    \label{fig:math_model}
\end{figure}

From our preceding assumptions, our model amounts to the following system on $\Omega \times (0,\infty)$:
\begin{align}
P_{,t} &= \overbrace{D_P \Lap P}^\text{diffusion} - \overbrace{c PS}^{\text{recruitment loss}} - \overbrace{f P}^\text{spontaneous misfolding} - \overbrace{a P,}^\text{breakdown}  \\
S_{,t} &= \overbrace{D_S \Lap S}^\text{diffusion} + \overbrace{c PS}^{\text{recruitment gain}} + \overbrace{f P}^\text{spontaneous misfolding} - \overbrace{b S,}^\text{breakdown} \end{align}
with
\begin{align}
P, S &\to 0, \quad |x| \to \infty, \\
-D_P \nabla P(x,t) \cdot \hat n &= \frac{-A}{1+(\frac{\langle S(\cdot,t-t_d)\rangle}{S_c})^m}, \quad x \in \partial \Omega, t \geq 0 \label{eq:Adim} \\
-D_S \nabla S \cdot \hat n &= 0, \quad \text{ on } \partial \Omega, \\ 
(P(x,0), S(x,0)) &= (P_0(x), S_0(x)), \quad x \in \bar \Omega, \\
S(x,t) &= S_-(x,t), \quad x \in \partial \Omega, t \in [-t_d,0]. 
\end{align}
Note that $\hat n$, the outward unit normal is based on the frame of $\Omega$. Hence, for the influx, we have $-A$ in Eq. \ref{eq:Adim}. We assume that $S_0(x) = S_-(x,0)$ on $\partial \Omega$. We prescribe that $P_0$ and $S_0$ satisfy the proper boundary conditions, namely $P_0, S_0 \to 0$ as $|x| \to \infty$, $\nabla S_0 \cdot \hat n = 0$ on $\partial \Omega$, and $-D_P \nabla P_0(x) \cdot \hat n = \frac{-A}{1+ (\frac{S_-(x,-t_d)}{S_c})^m}$ on $\partial \Omega.$ We note this is likely not strictly necessary as equations with diffusion tend to smooth out irregularities instantaneously. We also assume a finite total amount of protein in the system so that   $\int_{\Omega} P_0(x) \dd x, \int_{\Omega} S_0(x) \dd x < \infty.$ See Table \ref{tab:allparams} for a listing of all dimensional parameters. Figure \ref{fig:math_model} depicts this system.

\begin{table}
\centering
\begin{tabular}{c|c|p{2.in}} 
 Parameter & SI Base Units & Interpretation \\
 \hline
 \hline
  $D_P$ & $L^2/T$ & diffusion coefficient of P \\
  \hline 
  $D_S$ & $L^2/T$ & diffusion coefficient of S \\ 
  \hline 
  $c$ & $L^3/(MT)$ & rate of combination of P and S \\ 
  \hline 
  $f$ & $1/T$ & spontaneous misfolding rate of P \\ 
  \hline 
  $a$ & $1/T$ & clearance rate of P \\ 
  \hline 
  $b$ & $1/T$ & clearance rate of S \\ 
  \hline 
  $t_d$ & $T$ & delay time of the UPR \\ 
  \hline 
  $A$ & $M/(L^2 T)$ & maximum possible P-flux \\
  \hline 
  $S_c$ & $M/L^3$ & critical S-concentration \\
  \hline
  $R$ & $L$ & representative neuron size/radius \\
  \hline
 \end{tabular}
\caption{All dimensional parameters in the model where $M$ = mass, $L$ = length, and $T=$ time.} \label{tab:allparams}
\end{table}

Since a sustained UPR mechanism itself may cause damage to the cells \cite{moreno2012sustained,di2022disturbance}, we are interested in coming up with 
a proxy for what fraction of time the mechanism is active and to what extent. To that end, given the P-flux magnitude, $J$, we note that the quantity \begin{equation}h = 1-J/A \in [0,1) \label{eq:h} \end{equation} effectively describes whether the UPR is on or off at any given time and the closer $h$ is to $1$, the more strongly the UPR is activated. Then, the time average $h_\text{avg}$, gives a loose estimate for the fraction of time the UPR is active --- an effective UPR intensity. 

An alternative point of study would be to consider the frequency of UPR activations over times on $(0,T)$. We define 
\begin{equation}
W_T = \frac{1}{T} \big| \{ t | t \text{ is a strict local maximum for } h \text{ on } (0,T) \} \big|,
\end{equation}
which computes an average frequency of $h$-peaks on $(0,T).$

\begin{table}
\centering
\begin{tabular}{c|c|c|p{2.in}} 
 Scale & SI Base Units & Definition & Interpretation \\
 \hline
 \hline
 $\bar t$ & $T$ & $1/a$ & characteristic time for P to be cleared \\
 \hline 
 $\bar x$ & $L$ & $\sqrt{D_p/a}$ & characteristic distance P diffuses before being cleared \\
 \hline 
 $\bar P$ & $M/L^3$ & $\frac{A}{\sqrt{D_P a}}$ & characteristic scale of P as measured on the diffusive scale $\bar x,$ given the flux \\
 \hline 
 $\bar S$ & $M/L^3$ & $S_c$ & critical S-concentration to trigger decreased P-production \\
  \hline
 \end{tabular}
\caption{Scales chosen where $M$ = mass, $L$ = length, and $T=$ time.} \label{tab:scales}
\end{table}

\subsubsection{Nondimensionalization}

Nondimensionalizing and rescaling our system via $x = \bar x \tilde x$, $\Omega = \bar x \tilde \Omega$, $t = \bar t \tilde t$, $P(x,t) = \bar P p(\tilde x, \tilde t)$, and $S(x,t) = \bar S s(\tilde x, \tilde t)$ (see Table \ref{tab:scales}), after simplifying and removing tildes, on $\Omega \times (0, \infty)$, we have 
\begin{align}
    p_{,t} &= \Lap p - \gamma ps - (1+\sigma) p \label{eq:p}\\
    s_{,t} &= \delta \Lap s + \eta \gamma ps + \eta \sigma p - \beta s \label{eq:s} 
\end{align}
with
\begin{align}
    p, s &\to 0, \quad |x| \to \infty \label{eq:right} \\
    -\nabla p \cdot \hat n &= \frac{-1}{1+ \langle s(\cdot,t-\tau) \rangle^m} \quad x \in \partial \Omega, t \geq 0 \label{eq:pleft} \\
    -\nabla s \cdot \hat n &= 0, \quad \text{ on } \partial \Omega \label{eq:sleft} \\
    (p(x,0), s(x,0)) &= (p_0^*(x), s_0^*(x)), \quad x \in \bar \Omega \label{eq:ps_ic} \\
    s(x,t) &= s_-^*(x,t), \quad x \in \partial \Omega, t \in [-\tau,0] \label{eq:delay_ic}    
\end{align}
where $p_0^*$ and $s_0^*$ satisfy the correct boundary conditions at $t=0$ and $s_0(x) = s_-^*(x,0)$ on $\partial \Omega.$ The definitions of the dimensionless parameters are found in Table \ref{tab:ndim_params}.

\begin{table}
\centering
\begin{tabular}{c|c|p{2.in}} 
 Parameter & Definition & Interpretation\\
 \hline
 \hline 
 $\gamma$ & $\frac{c S_c}{a}$ & ratio of (the rate P is converted to S when S concentration is $S_c$) to (the rate P is cleared)  \\
 \hline 
 $\sigma \ll 1$ & $f/a$ & ratio of the rate P misfolds to the rate it is cleared  \\
 \hline 
 $\delta \leq 1$ & $D_S/D_P$ & ratio of S diffusivity to P diffusivity \\
 \hline 
 $\eta$ & $\frac{A}{S_c \sqrt{D_P a}}$ & ratio of characteristic P-concentration to $S_c$ \\
 \hline 
 $\beta \leq 1$ & $b/a$ & ratio of rate S is cleared to the rate P is cleared \\
 \hline 
 $\tau$ & $a t_d$ & delay relative to P-clearance time \\
  \hline 
  ${\rho}$ & $\frac{R \sqrt{a}}{\sqrt{D_P}}$ & ratio of (the characteristic cell size) to (the characteristic distance P travels before being cleared) \\
  \hline
 $m>0$ & chosen value & controls abruptness of switch
\end{tabular}
\caption{Dimensionless parameters appearing in the model.} \label{tab:ndim_params}
\end{table}

Given the diffusive nature of the system, we anticipate smooth solutions to the system of Equations \eqref{eq:p}-\eqref{eq:delay_ic} that exist globally in time. The proof of existence of such solutions and the properties they enjoy is left as future work. Hereafter, we assume the existence of classical solutions $p, s \in \mathcal{C}^2_x \cap \mathcal{C}^1_t (\Omega, [0,\infty)).$

Note that through nondimensionalization and rescaling, we reduce a system with $10$ dimensional parameters (Table \ref{tab:allparams}) to a dimensionless system involving $7$ dimensionless parameters (Table \ref{tab:ndim_params}), excluding $m$. It is useful to consider that all dimensional systems resulting in the same set of dimensionless parameters have the same dimensionless solutions. 

\subsubsection{Simplified Geometries}

\begin{figure}
\includegraphics[width=0.9\linewidth]{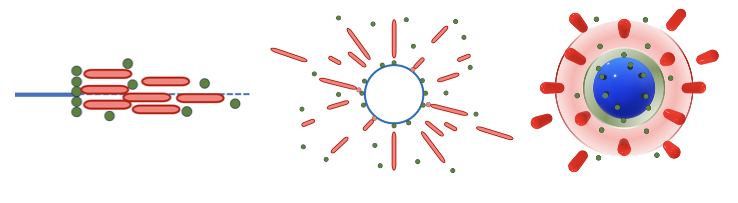}
    \caption{Reductions to $1-$dimensional (left), $2-$dimensional radially symmetric (middle), and $3-$dimensional radially symmetric (right) geometries, respectively. The green dots represent the P proteins, while the red rods represent the S proteins.   } \label{fig:symmetry}
\end{figure}

As specific scenarios, we consider $d=1$ in a planar geometry and $d=2,3$ in radially symmetric geometries --- effectively describing an idealized ``spherical neuron" in different dimensions. See Figure \ref{fig:symmetry}. We denote $r=|x|$ and reduce $\Omega$ to $B = (\rho,\infty).$ Then 
$$\int_\Omega z(x) \dd x = C_d \int_{\rho}^\infty z(r) r^{d-1} \dd r$$
where $$C_d = \begin{cases} 1, \quad d= 1, \\
2 \pi, \quad d=2, \\
4 \pi, \quad d=3 
\end{cases}$$ is a surface area factor.

In the symmetric geometry, on $B \times (0, \infty)$, we have
\begin{align}
    p_{,t} &= p_{,rr} + \frac{d-1}{r} p_{,r} - \gamma ps - (1+\sigma) p, \label{eq:pr}\\
    s_{,t} &= \delta ( s_{,rr} + \frac{d-1}{r} s_{,r} ) + \eta \gamma ps + \eta \sigma p - \beta s, \label{eq:sr} 
\end{align}
with
\begin{align}
    p, s &\to 0, \quad r \to \infty, \label{eq:right} \\
    p_{,r}|_{r={\rho}} &= \frac{-1}{1+ s({\rho},t-\tau)^m} \quad t \geq 0, \label{eq:pleftr} \\
    s_{,r}|_{r={\rho}} &= 0, \label{eq:sleftr} \\
    (p(r,0), s(r,0)) &= (p_0(r), s_0(r)), \quad r \in \bar B, \label{eq:ps_icr} \\
    s({\rho},t) &= s_-(t), \quad t \in [-\tau,0], \label{eq:delay_icr}\\
    h_\text{avg} &= \lim_{\Upsilon \to \infty} \frac{1}{\Upsilon} \int_0^\Upsilon (1 - \frac{1}{1+ s({\rho},t-\tau)^m}) \dd t, \label{eq:damage} \\
    \omega_\text{avg} &= \lim_{\Upsilon \to \infty} \frac{1}{\Upsilon} \times \nonumber \\ &\big| \{ t | t \text{ is a strict local maximum for } s(\rho,t-\tau) \text{ on } (0, \Upsilon) \} \big|, \label{eq:damage_om}
\end{align}
where $p_0$ and $s_0$ satisfy the correct boundary conditions at $t=0$, and where $s_-(0) = s_0({\rho}).$ We have added the formulas for the average fraction of time the UPR is activated, Eq. \eqref{eq:damage}, and the average dimensionless frequency of UPR activations, Eq. \eqref{eq:damage_om}. We shall keep $m=10$ fixed throughout this work as it allows for a relatively rapid switch term.

We choose initial conditions to model the introduction of a highly localized quantity of S at position $r=r_* > {\rho}$ so that initially $s\approx 0$ is negligibly small at $r={\rho}.$ We let \begin{equation} i_p := C_d \int_{\rho}^\infty r^{d-1} p_0(r) \dd r \label{eq:ip} \end{equation} and \begin{equation} i_s := C_d \int_{\rho}^\infty r^{d-1} s_0(r) \dd r \label{eq:is} \end{equation} represent the initial quantity of the proteins and \begin{equation} -p_{0,r}|_{r={\rho}} = n_p \label{eq:np} \end{equation} specify the normal derivative of $p_0$ at the cell boundary where $s_0=0$. We then choose
\begin{align}
p_0(r) &= c_{1d} r^{1-d} \e^{-c_{2d} (r-{\rho}) - (r-{\rho})^2}, \label{eq:pic} \\
s_0(r) &= \frac{i_s r^{1-d}}{C_d \omega \left( 1 + \tanh(\frac{r_*-{\rho}}{\omega}) \right)} \mathrm{sech}^2(\frac{r-r_*}{\omega}), \label{eq:sic} \\
s_-(t) &=  s_0(\rho), \quad t \in [-\tau,0], \label{eq:pastic}
\end{align}
where $c_{1d}$ and $c_{2d}$ satisfy the system
\begin{align*}
    \frac{\sqrt{\pi}c_{1d}}{2} \e^{c_{2d}^2/4} \mathrm{erfc}(c_{2d}/2) &= i_p/C_d, \\ 
    c_{1d} c_{2d} {\rho}^{1-d} + (d-1) c_{1d} c_{2d}^{-d} &= n_p,
\end{align*}
and $\omega=0.2$ is a shape parameter. Under this setup, $C_d \int_{\rho}^\infty r^{d-1} s_0(r) \dd r = i_s.$ With $\omega$ small, $s_0$ and its derivatives are very small for $r_*-r = O(1)$ so that $s_0({\rho}), s_0'({\rho}) \approx 0$ are negligibly small. Then with $s_0({\rho}) = 0,$ boundary conditions require that $p_0'({\rho}) = -1.$ The constants $c_{1d}$ and $c_{2d}$ ensure $p_0'({\rho}) = -n_p$ and $C_d \int_{\rho}^\infty r^{d-1} p_0(r) \dd r = i_p.$ For most problems, $i_p = i_s = n_p = 1$, but this system is more general.

We make one final remark on the total protein quantity. In $d=1$, we interpret the system as describing protein concentrations in a very narrow, straight region. The cell membrane is one boundary and concentration variations orthogonal to the direction pointing away from the cell are negligible. Thus, the system has a small cross-sectional area and is effectively one-dimensional. Likewise, in $d=2$, we interpret the system as describing protein concentrations when primarily confined to a plane with limited variation normal to the plane, with concentration varying with distance from a circular cell membrane. The system has a narrow height and is primarily two dimensional with the cell membrane being a circle. In $d=3$, we interpret the system as concentrations of proteins as their radial distance from the centre of a sphere varies. Thus, for $d=1$, $i_p$ is expressing a dimensionless total quantity of P per unit area of the region. Likewise with $d=2$, $i_p$ is the total dimensionless quantity of P per unit height of the region. Then with $d=3$, $i_p$ is the total dimensionless quantity of P. The same applies to $i_s$.

\subsection{Numerical Method}

\label{sec:numerical}

Owing to the challenging nature of the equations, we analyze our model numerically. Full details can be found in the Appendix \ref{app:num}. In brief: we implement a finite difference scheme that is first order in time, second order in space, with semi-implicit timestepping.

\section{Math and Model Investigations}

\label{sec:results}

This section is dedicated to studying the qualitative behavior of the model itself over a range of parameters. Numerical specifications are given in Appendix \ref{app:model}.

\subsection{Model Simulations}

\begin{figure}
    \centering
    \includegraphics[width=1.\linewidth]{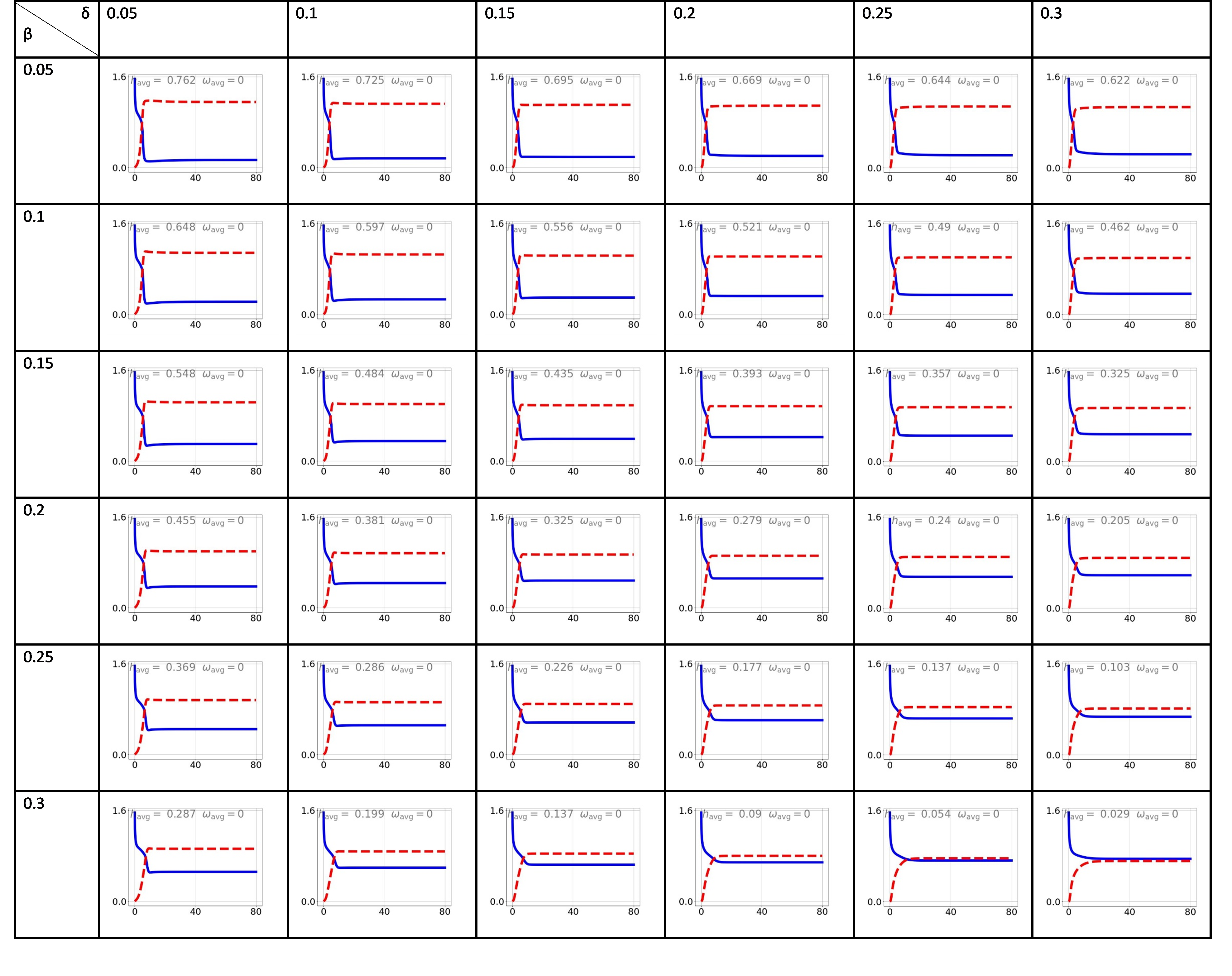}
    \caption{Phase diagram at $\tau = 0$ with $d=1$ along membrane $r=\rho$. The horizontal axis is $t$, the vertical axis is concentration with $p$ solid blue and $s$ dashed red. Unspecified parameter values are $\gamma=1,$ $\sigma=0.02,$ $\eta=1$, and $\rho=0.25$.} \label{fig:phase0}
\end{figure}

\begin{figure}
    \centering
    \includegraphics[width=1.\linewidth]{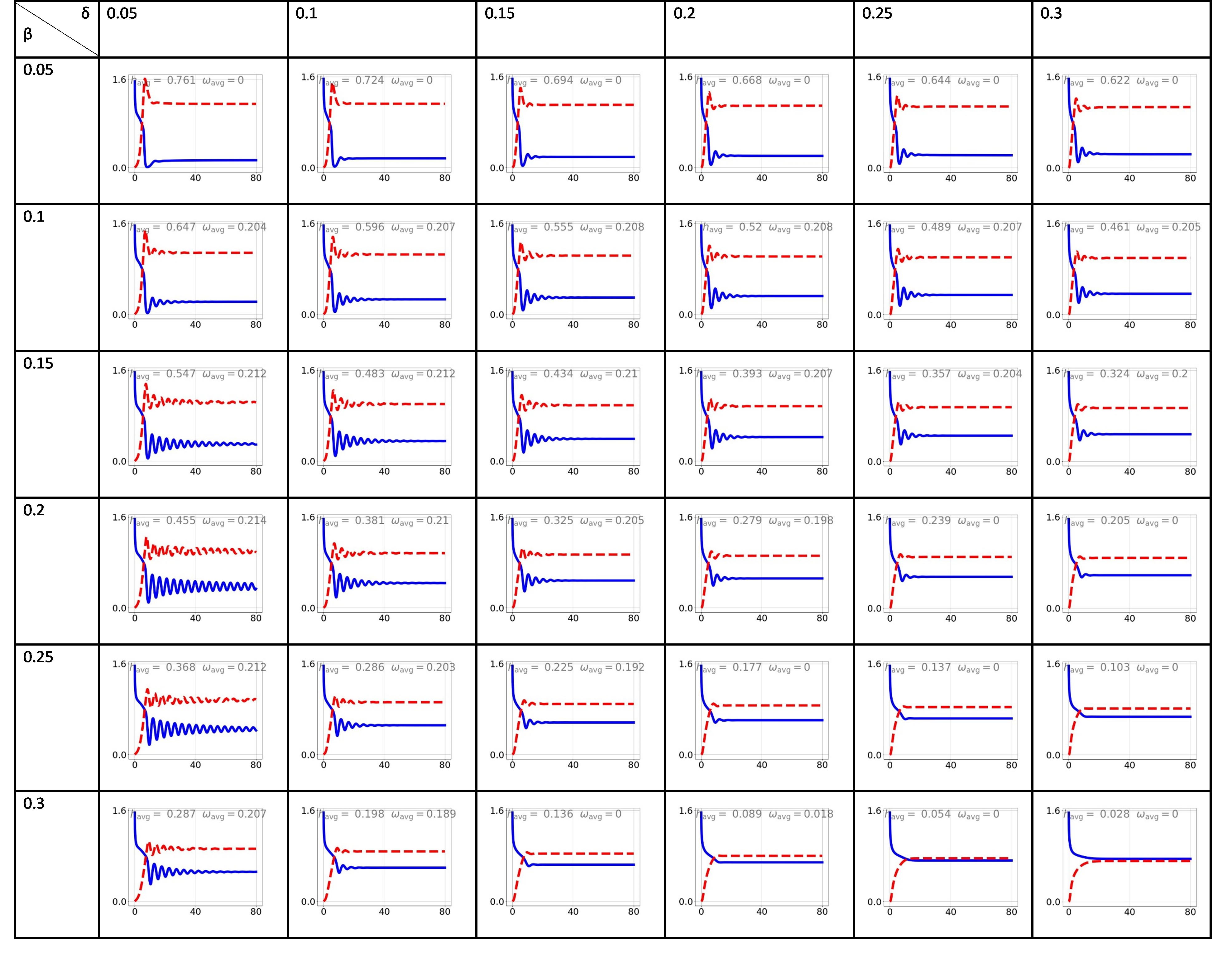}
    \caption{Phase diagram at $\tau = 1$ with $d=1$ along membrane $r=\rho$. The horizontal axis is $t$, the vertical axis is concentration with $p$ solid blue and $s$ dashed red. Unspecified parameter values are $\gamma=1,$ $\sigma=0.02,$ $\eta=1$, and $\rho=0.25$.} \label{fig:phase1}
\end{figure}

\begin{figure}
    \centering
    \includegraphics[width=1.\linewidth]{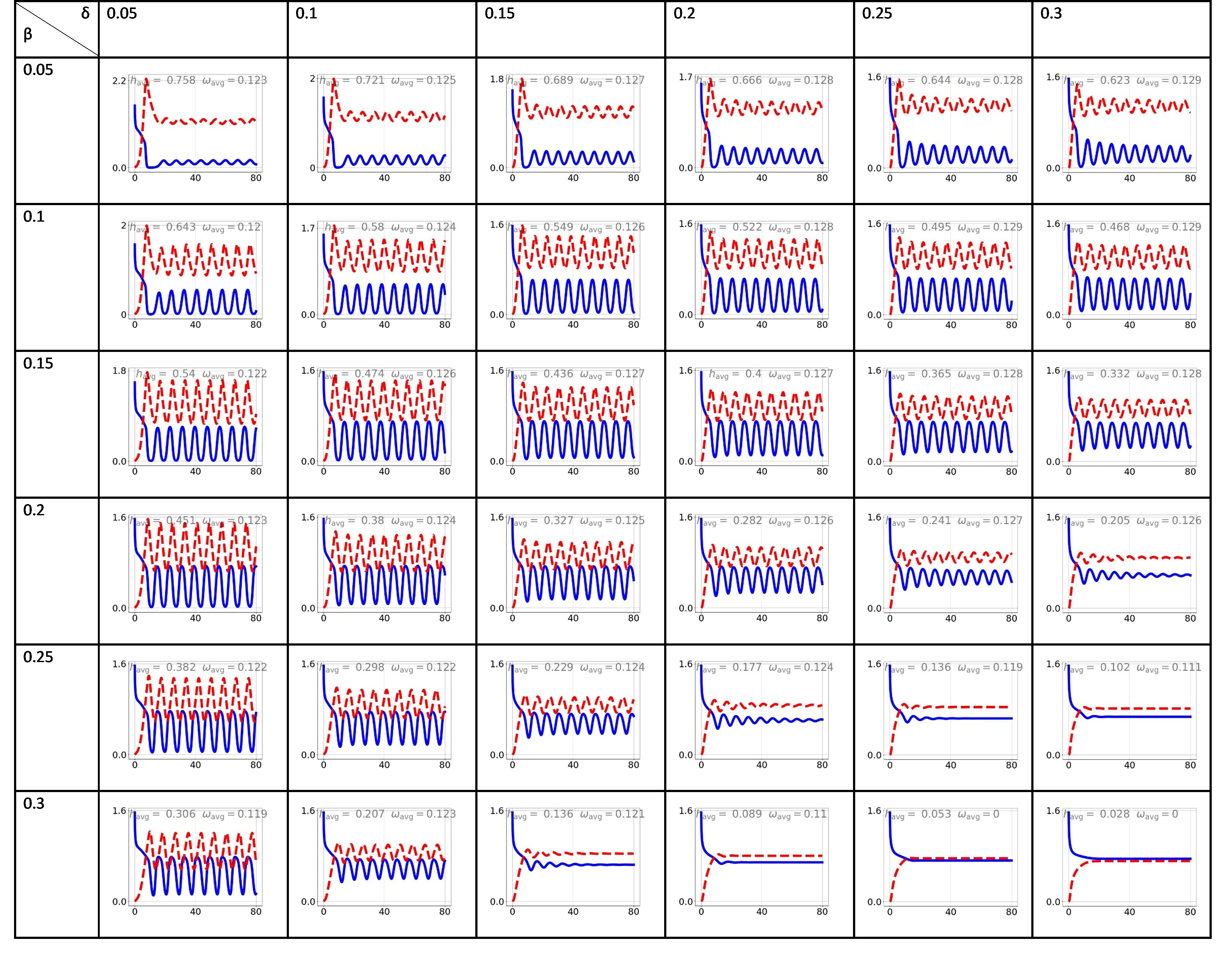}
    \caption{Phase diagram at $\tau = 2$ with $d=1$ along membrane $r=\rho$. The horizontal axis is $t$, the vertical axis is concentration with $p$ solid blue and $s$ dashed red. Unspecified parameter values are $\gamma=1,$ $\sigma=0.02,$ $\eta=1$, and $\rho=0.25$.} \label{fig:phase2}
\end{figure}

\begin{figure}
    \centering
    \includegraphics[width=1.\linewidth]{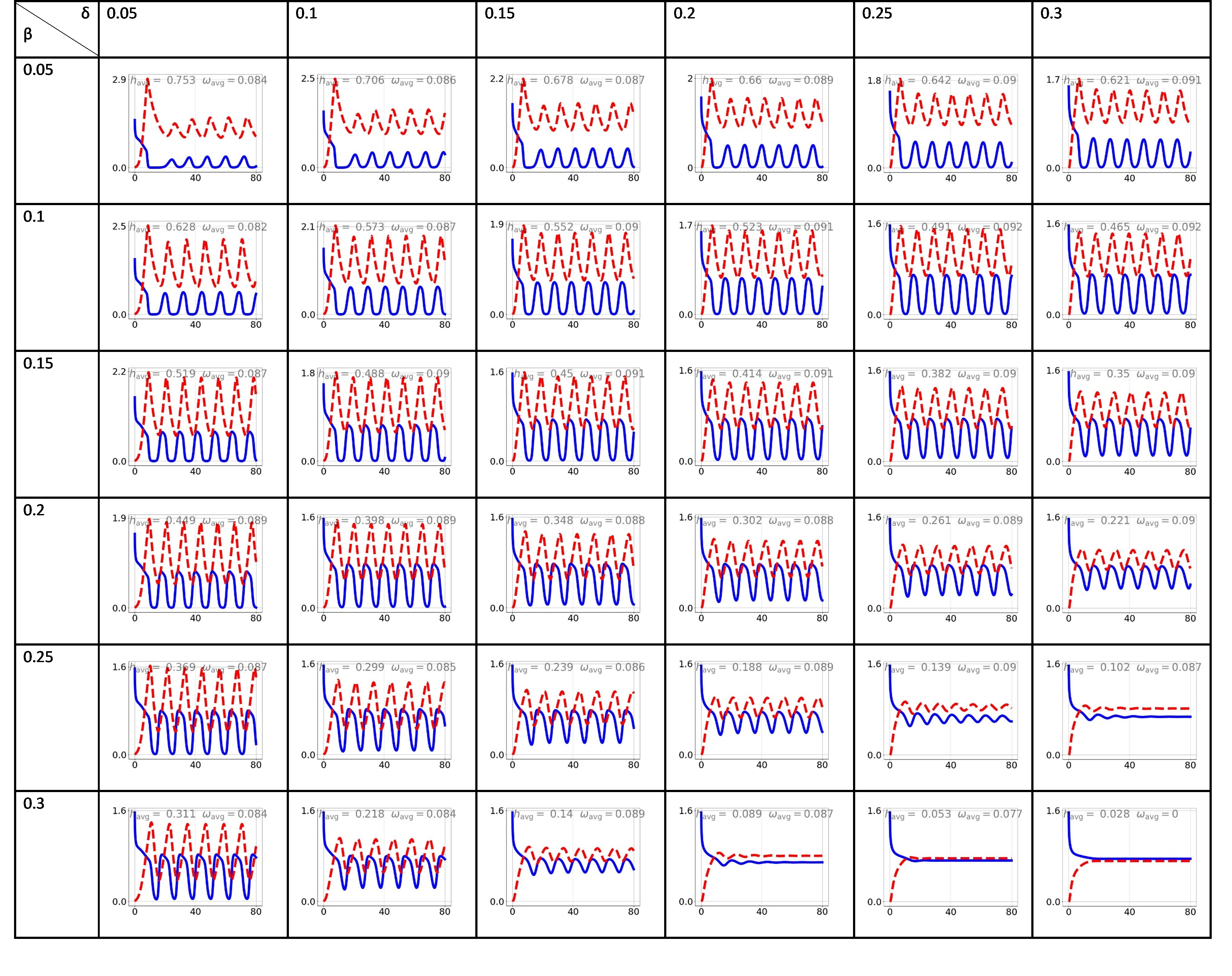}
    \caption{Phase diagram at $\tau = 3$ with $d=1$ along membrane $r=\rho$. The horizontal axis is $t$, the vertical axis is concentration with $p$ solid blue and $s$ dashed red. Unspecified parameter values are $\gamma=1,$ $\sigma=0.02,$ $\eta=1$, and $\rho=0.25$.} \label{fig:phase3}
\end{figure}

\begin{figure}
    \centering
    \includegraphics[width=1.\linewidth]{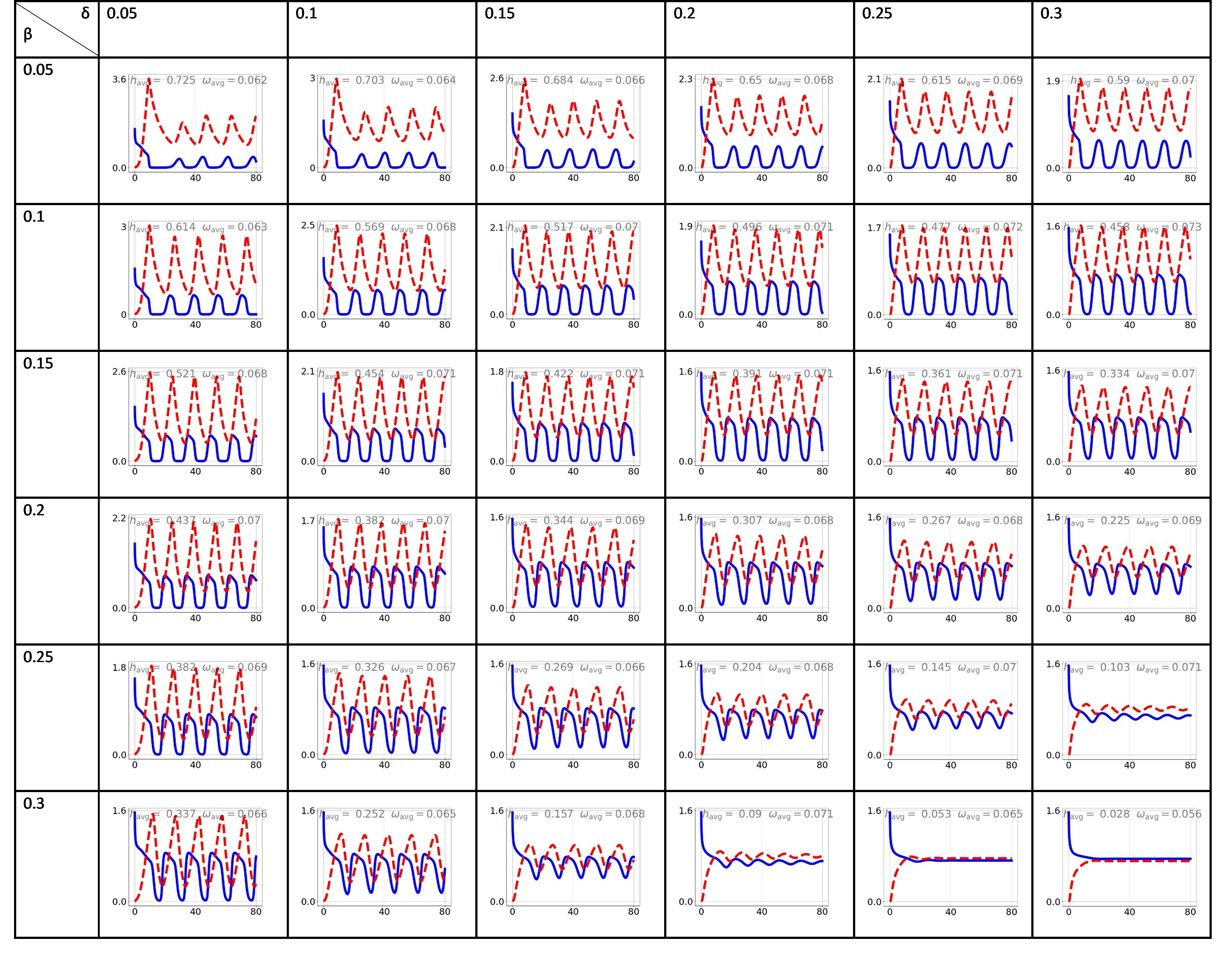}
    \caption{Phase diagram at $\tau = 4$ with $d=1$ along membrane $r=\rho$. The horizontal axis is $t$, the vertical axis is concentration with $p$ solid blue and $s$ dashed red. Unspecified parameter values are $\gamma=1,$ $\sigma=0.02,$ $\eta=1$, and $\rho=0.25$.} \label{fig:phase4}
\end{figure}

\begin{figure}
    \centering
    \includegraphics[width=1.\linewidth]{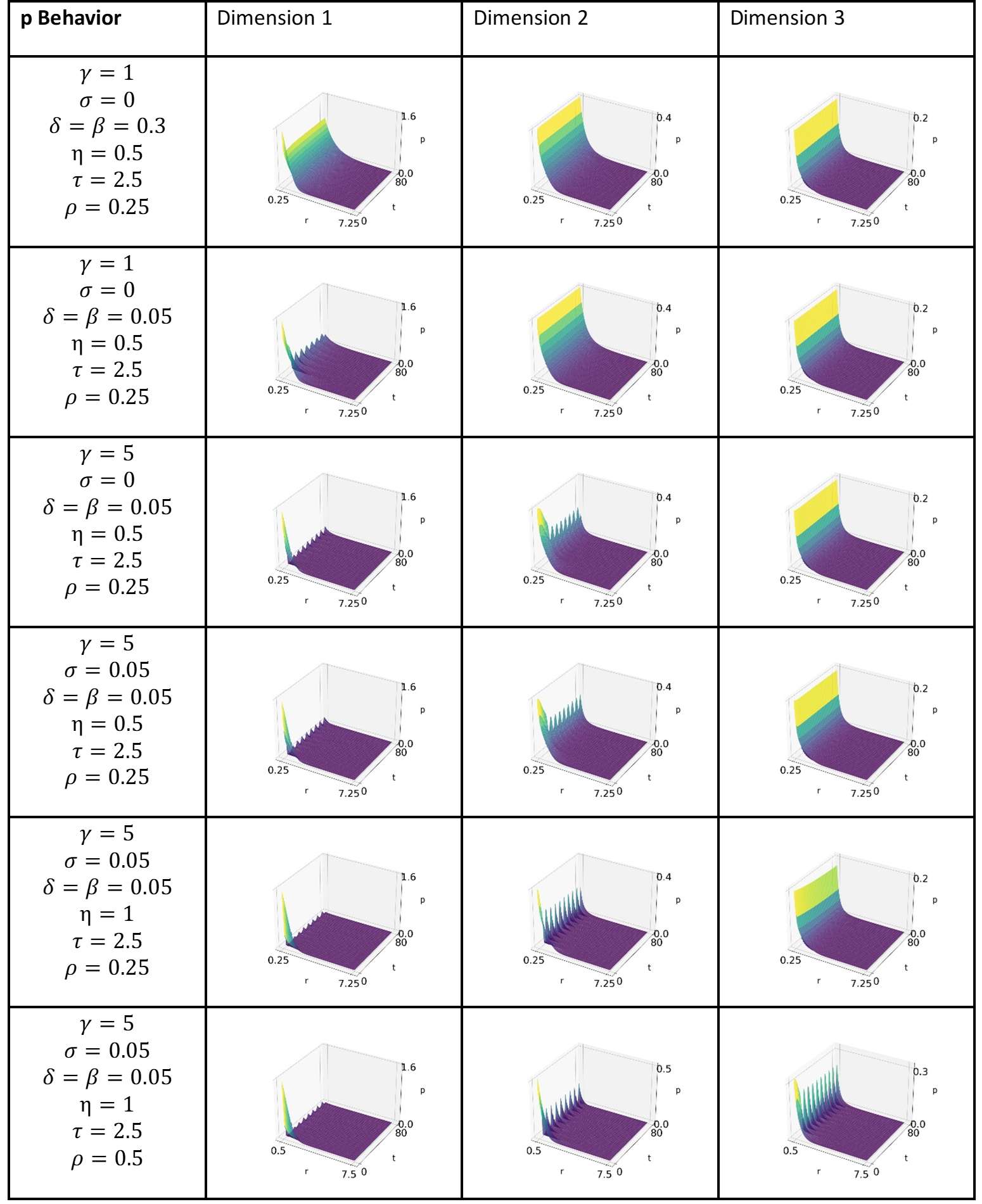}
    \caption{Behavior of $p$ with varying parameters.}  
    \label{fig:pbehavior}
\end{figure}

\begin{figure}
    \centering
    \includegraphics[width=1.\linewidth]{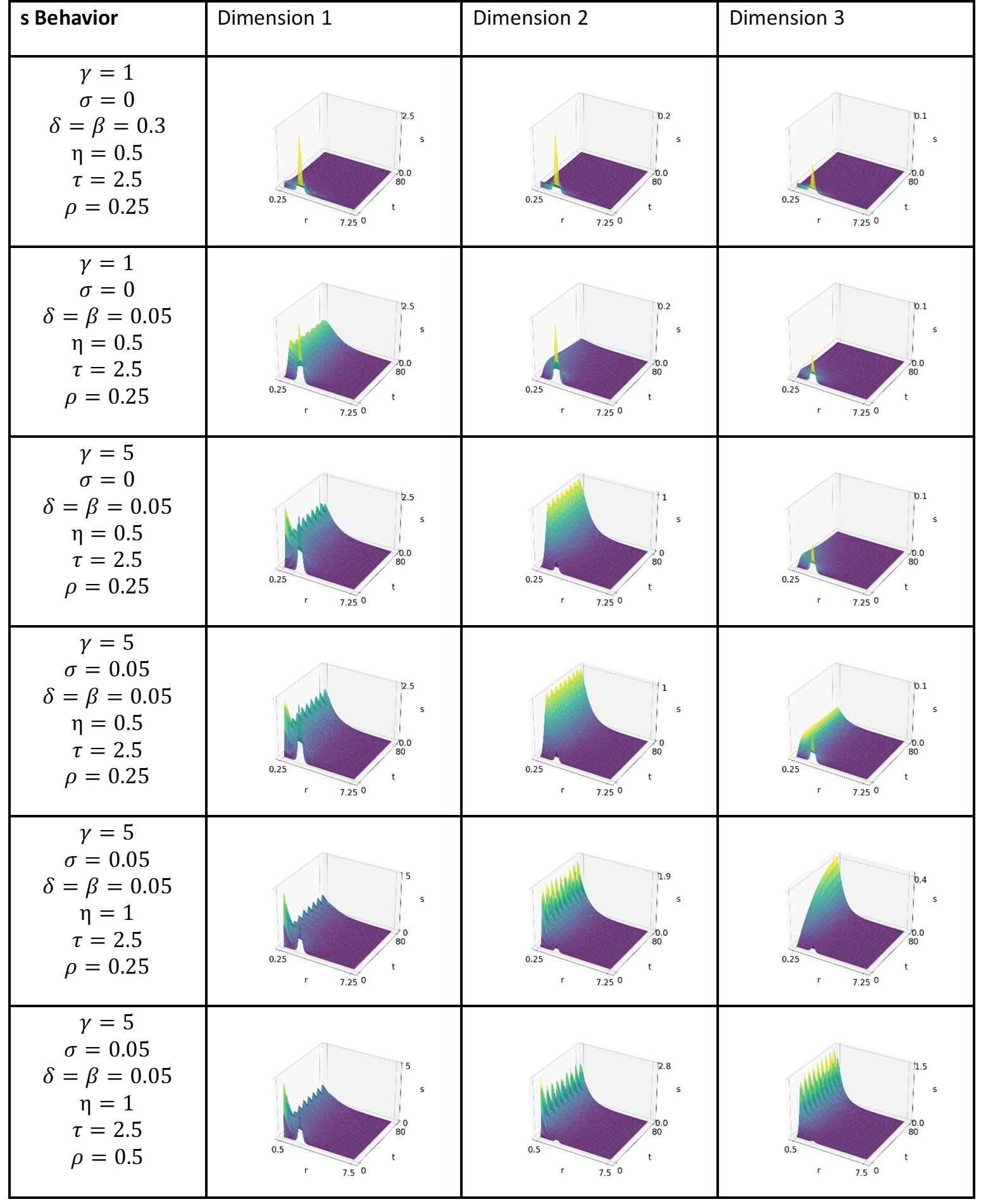}
    \caption{Behavior of $s$ with varying parameters.} \label{fig:sbehavior}
\end{figure}

To study how the system responds to different parameters, we begin by considering variations in $\beta$ and $\delta$ at different fixed values of $\tau$. In making these explorations, we focus primarily on phase portraits for $d=1$ dimensions depicting the protein concentrations at the cell membrane over time (since this is the driving mechanism of the UPR). This is found in Figures \ref{fig:phase0}-\ref{fig:phase4}. We also consider the spatiotemporal spreading of the proteins with further experiments in Figures \ref{fig:pbehavior} and \ref{fig:sbehavior}. We wish to caution the reader that the ranges of values along the vertical differ between trials.

From the phase portraits, Figures \ref{fig:phase0}-\ref{fig:phase4}, we observe that oscillations tend to decrease in intensity and are dampened out as $\beta$ increases, as $\delta$ increases, and as $\tau$ decreases. The UPR intensity $h_\text{avg}$ tends to decrease as $\beta$ and $\delta$ increase, and does not vary significantly as $\tau$ varies in the experiments. The UPR frequency $\omega_\text{avg}$, when present, is quite consistent in its values across the phase portraits for fixed $\tau$-values, and tends to decrease as $\tau$ increases.

From the spatiotemporal plots for parameter variations, Figures \ref{fig:pbehavior}-\ref{fig:sbehavior}, we observe that the oscillations become less pronounced as the dimension $d$ increases. We again observe how decreasing $\delta$ and $\beta$ can make the system oscillatory. Increasing $\gamma$ makes these oscillations more pronounced. The effect of $\sigma$ on the oscillations is quite small. Increasing $\eta$ makes the oscillations more pronounced in the $d=2$ case and drastically increases the $s$-values in $d=3.$ Finally, when $\rho$ is increased, the $d=2$ system becomes more oscillatory and the $d=3$ system begins to oscillate (the only set of oscillations observed in the spatiotemporal plots). The fact that $\rho$ has no effect in $d=1$ is unsurprising since there is no geometric $\frac{d-1}{r} \partial_r$ term appearing in the Laplacian. As $\rho$ increases, the $d=2$ and $d=3$ systems are more planar on the cell membrane, so it makes sense that with $d=1$ being highly oscillatory in general, increasing $\rho$ should make $d=2$ and $d=3$ more oscillatory.

\begin{figure}
    \centering
    \includegraphics[width=0.3\linewidth]{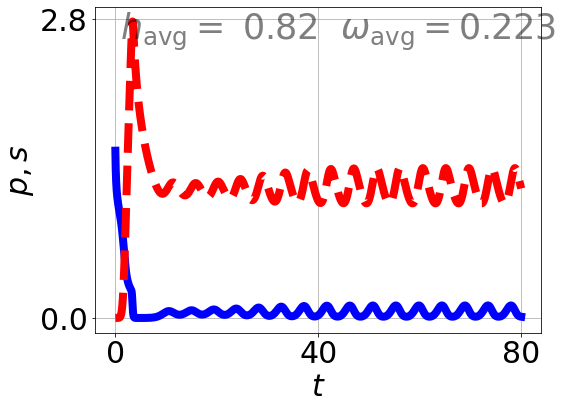} \includegraphics[width=0.3\linewidth]{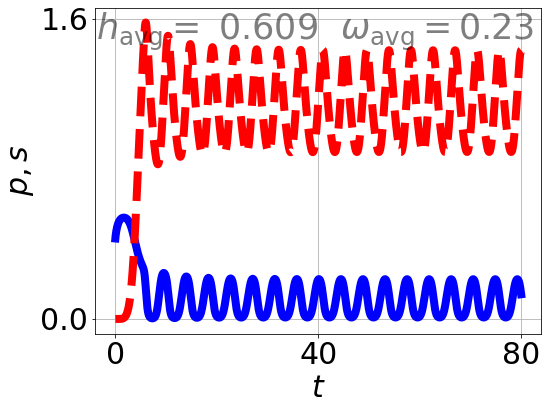} \includegraphics[width=0.3\linewidth]{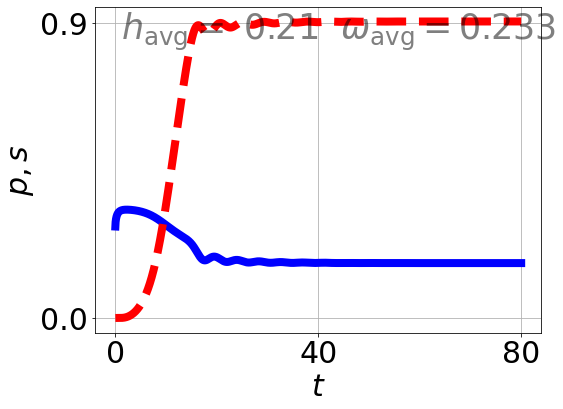} 
    \caption{Temporal variations in concentration at $r=\rho$ in $d=1$ (left), $d=2$ (middle), and $d=3$ (right) with $p$ solid blue and $s$ dashed red. Parameters are $\gamma=6$, $\sigma=0$, $\delta=\beta=0.075$, $\eta=0.9$, $\tau=1.25$, and $\rho=0.5$.} \label{fig:extra}
\end{figure}

In Figure \ref{fig:extra}, we illustrate how the oscillations and UPR intensity are weakened as the dimension of the system $d$ increases.

\subsection{Comparison with Previous Work}

Having explored many of the parameters in our model, we briefly turn to the question of how this new model compares with the DDE model of Adimy \cite{adimy2022neuron} which inspired this work. The DDE model also studied a heterodimer system with a production rate for healthy monomers with delay, clearance rates for both the healthy and toxic proteins, and a recruitment rate to convert healthy proteins to toxic proteins. The authors found that for low levels of delay, the system could be stable (without oscillations) and that a bifurcation could take place by increasing the delay to yield an oscillatory solution. With the DDEs, a locally and globally asymptotically stable disease-free steady state was found when the reproductive ratio $\mathcal{R}_0 < 1$, where $\mathcal{R}_0$ gives the ratio of the product of the peak production rate and recruitment rate to the product of the clearance rates. When $\mathcal{R}_0 > 1$, the disease takes overand oscillatory solutions may occur.

\begin{figure}
\includegraphics[width=0.23\linewidth]{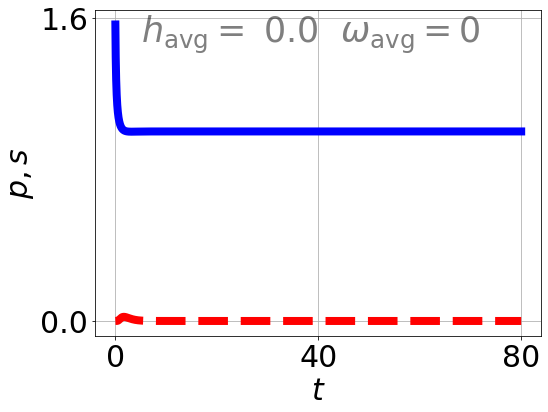} \includegraphics[width=0.23\linewidth]{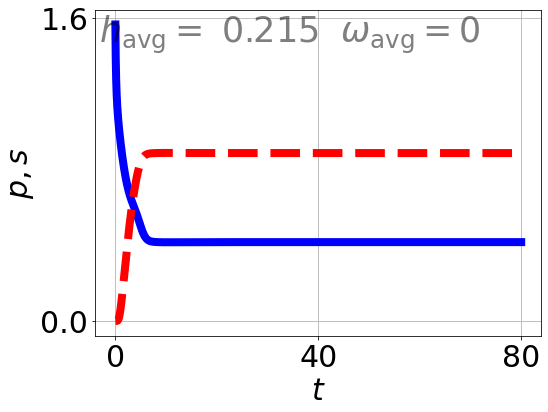} 
\includegraphics[width=0.23\linewidth]{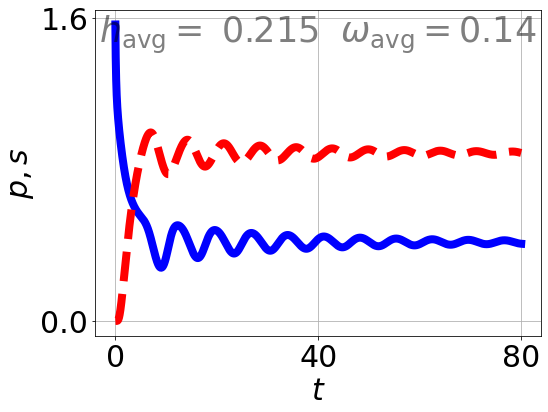} \includegraphics[width=0.23\linewidth]{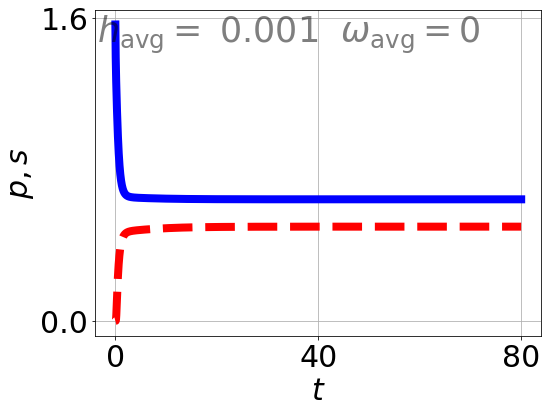} 
\caption{Temporal variations in concentration at $r=\rho$ with $d=1$ with $p$ solid blue and $s$ dashed red. From left to right: disease-free ($\gamma=0.2,$ $\sigma=0,$ $\delta=0.2$, $\eta=0.5$, $\beta=1$, $\tau=0.2$, and $\rho=0.25$); endemic, no oscillations ( $\gamma=3,$ $\sigma=0,$ $\delta=0.2$, $\eta=0.5$, $\beta=0.2$, $\tau=0.2$, and $\rho=0.25$ ); endemic, oscillations ( $\gamma=3,$ $\sigma=0,$ $\delta=0.2$, $\eta=0.5$, $\beta=0.2$, $\tau=2$, and $\rho=0.25$ ); and endemic, no oscillations ( $\gamma=3,$ $\sigma=0,$ $\delta=1$, $\eta=0.5$, $\beta=0.2$, $\tau=2$, and $\rho=0.25$  ).} \label{fig:compare}

\end{figure}

Here, we illustrate how our system observes similar behavior. In Figure \ref{fig:compare}, we begin with parameters for a disease-free state with small $\gamma$ and large $\beta$, bring about a disease state with increasing $\gamma$, induce oscillations with an increase in $\tau$, and then eliminate the oscillations with an increase in $\delta.$ We note that while the behavior is similar to the DDE model, factors like $\delta$ stemming from diffusion also play a role. And as noted in Figures \ref{fig:pbehavior}-\ref{fig:extra}, geometric factors such as the cell size $\rho$ and the dimension $d$ of the spatial domain are relevant.

\subsection{Mathematical Perspective}

Given the nonlinear structure in the model, especially within the delayed flux boundary condition for P, there are open questions as to the well-posedness of the PDE system and to the solution properties. For example, a function in $g \in \mathcal{C}^2_x\mathcal{C}^1_t(\Omega; (0,\infty))$ could exhibit unbounded behavior for $x$ near $\partial \Omega$ as $t \to \infty$ while retaining a bounded normal derivative. We expect solutions with intuitive, physical bounds to exist, but this is left as a future work in mathematics. Additionally, while delay differential equations are well-studied, having a delay in the boundary condition itself is somewhat unique.

We additionally observe several ``bifurcations" that emerge in our system, such as the turning off/on of oscillations as model parameters vary or going from a disease-free to diseased state. We have 7 dimensionless parameters (excluding $m$) and it would be of great interest to understand their individual effects from a rigorous, theoretical perspective. We also note that, due to the limitations of a numerical study, some of the observed oscillations that appear to be damped  may actually be sustained at a small amplitude in the limit, and some of the systems that appear to have no oscillations may actually oscillate. This makes further mathematical study important.

\section{Biological Implications}

\label{sec:bio}

We now focus our attention on the biological and medical insights provided by the model.

\subsection{Parameter Estimation and Setup}

To our knowledge, extant literature does not provide data to accurately determine the many parameters used in our model. It would be of interest to conduct experiments that directly measure the chemical reaction rates, diffusivities, etc., {\it in vivo}. Nevertheless, we venture to obtain {\it very loose estimates} for the parameters based on the literature to then model, as best possible, a representative biological system. The data we use come from a variety of animal models, tissue types, and mathematical models with different objectives than our own. Our parameters are listed in Table \ref{tab:param_vals} in cgs units and justifications follow in the paragraphs below. The resulting dimensionless parameters are then given in Table \ref{tab:ndim_vals}. We only utilize one significant figure to reflect the parameter uncertainty.

\paragraph{Parameters $D_P$ and $D_S$:} The cellular prion protein PrP$^\text{C}$ has a molecular mass of $27-30$ kDa \cite{martin1992applications}. We shall use $28.5$ kDa. Interstitial fluid has been reported to have a range of viscosities of $0.7-3.5$ cP \cite{bera2022extracellular} and we shall use the mean of the extremes for a value $2.1$ cP. A useful engineering approximation is that the diffusivity of a protein is given by \begin{equation} D \approx \frac{\zeta T}{v m^{1/3}} \label{eq:Dapp} \end{equation} where $\zeta = 8.34 \times 10^{-8} \frac{\text{cm}^2 \cdot \text{cP} \cdot \text{Da}^{1/3}}{\text{s} \cdot (^\circ \text{K})}$, $T$ is the absolute temperature, $v$ is the solvent viscosity, and $m$ is the molar mass of the solute \cite{young1980estimation}. However, due to tortuosity, this is scaled down further by $\iota^2$ where $\iota \approx 1.6$ is the tortuosity of the brain \cite{hrabe2004model}. At $T=310^\circ$K, this yields $D_P \approx 2 \times 10^{-7}$ cm$^2$/s.

In a nucleated polymerization model based on mouse brain data, the mean number of monomers in a PrP$^\text{Sc}$ protein is estimated at $100-10000$ \cite{masel1999quantifying}. Using the geometric mean, this suggests our S-proteins should have a mass that is $1000$ times larger than our P-proteins. Based on diffusivity scaling with the inverse cube root of molecular mass in Equation \eqref{eq:Dapp}, we find $D_S \approx 2 \times 10^{-8}$ cm$^2$/s.

\paragraph{Parameters $a$ and $b$:} Based on mouse and hamster brain data, a nucleated polymerization model estimated the clearance rate of PrP$^\text{C}$ at approximately $3-5$ day$^{-1}$ \cite{masel1999quantifying}. We use a value of $4$ day$^{-1}$ so $a \approx 5 \times 10^{-5}$ s$^{-1}$. The rate of clearance of toxic Scrapie proteins was also estimated to be $0.03-0.2$ day$^{-1}$ and we use the geometric mean of the extremes yielding $b\approx0.0775$ day$^{-1} \approx 9 \times 10^{-7}$ s$^{-1}.$

\paragraph{Parameter $S_c$:} For hamsters at the terminal stage of Scrapie, the concentration of Scrapie proteins was measured to be about $100$ $\mu$g per gram of brain tissue \cite{beekes1996sequential}. The density of brain tissue is approximately $1$ g/cm$^3$ \cite{barber1970density}. Taking the terminal concentration as the critical concentration $S_c$, we have $S_c= 100$ $\mu$g/cm$^3 =1 \times 10^{-4}$ g/cm$^3.$

\paragraph{Parameter $f$:} Due to the rarity of spontaneous prion diseases, we take $f=0$ s$^{-1}$.

\paragraph{Parameter $c$:} Based on hamster data, a heterodimer model estimates the combination rate at $0.15$ $\frac{\text{g brain}}{\mu \text{g} \cdot \text{day}}$ \cite{matthaus2006diffusion}. Again with a brain density of $1$ g/cm$^3$ \cite{barber1970density}, we find $c \approx 0.15$ $\frac{\text{cm}^3}{\mu \text{g} \cdot \text{day}} \approx 2 \frac{\text{cm}^3}{\text{g} \cdot \text{s}}.$

\paragraph{Parameter $t_d$:} A study that infected human fibrosarcoma cells with Dengue Virus found an increase in phosphorylated elF2$\alpha$ --- indicating the activation of Integrated Stress Response arm of the UPR --- at approximately $6$ hours post infection \cite{pena2011dengue}. At at $24-36$ hours in that same study, the inositol-requiring-protein-1 UPR pathway was activated. This suggests a timescale for the UPR may be in the loose range of $6-36$ hours, and we use the mean of the extremes, $t_d \approx 21$ hours or $t_d \approx 80000$ s.

\paragraph{Parameter $A$:}  The most challenging parameter to estimate is the maximum flux of P-proteins. We first note that a heterodimer model on hamster data estimates the rate of monomer production to be $\lambda = 4$ $\frac{\mu \text{g}}{\text{(g brain)} \cdot \text{day}}$ \cite{matthaus2006diffusion}, which with brain density of $1$ g/cm$^3$ \cite{barber1970density}, we have $\lambda \approx 5 \times 10^{-11} \frac{\text{g}}{\text{cm}^3 s}.$

There are approximately $2 \times 10^9$ neurons per cm$^3$ \cite{braitenberg2001brain} across many species. Neuronal perikaryal diameters range from about $6$ to $80$ $\mu$m and the perikarya accounts for approximately 10\% of the neuron surface area \cite{raine}. By these estimate and with the geometric mean of the diameter lower and upper bounds taken as a length scale, the surface area of a neuron is $\sim 5000$ $\mu$m$^2$ or $5 \times 10^{-5}$ cm$^2$.

Then, as a ballpark figure for the flux of P-proteins from a single neuron, we obtain $A^* \approx \frac{5 \times 10^{-11}}{ 2 \times 10^9 \times 5 \times 10^{-5}} = 5 \times 10^{-16}$ $\frac{\text{g}}{\text{cm}^2 \cdot \text{s}}.$ To this figure, we make a scaling correction. As our intent with the model is to understand the dynamics experienced by a single typical neuron, we need to take into account that no neuron functions in isolation and the brain is full of neurons. Neurons in a nearby vicinity will also experience similar stresses due to the presence of S-proteins. We thereby choose to scale $A^*$ by a ballpark estimate of the number of neurons that could influence any particular neuron through their release of P-proteins.

On average, a P-protein survives a time of $1/a$, resulting in a characteristic diffusive lengthscale of $\sqrt{D_p/a} \approx 0.06$ cm. Within a volume of $0.06^3$ cm$^3$, there are approximately $4 \times 10^5$ neurons, and we scale $A^*$ up by this to give a nominal flux of $A \approx 2 \times 10^{-10}$ $\frac{\text{g}}{\text{cm}^2 \cdot \text{s}}.$

\paragraph{Parameter $R$:} cells of the globus pallidus in humans can have dendritic extends of $\approx 1000$ $\mu$m and soma radii of $\approx 16.5$ $\mu$m \cite{stuart2016dendrites}. To come up with a representative length scale for an idealized ``spherical neuron", we take a geometric mean of these length scales giving $R \sim 128$ $\mu$m $\approx 0.01$ cm.

\begin{table}
\centering
\begin{tabular}{c|c|p{2.in}} 
 Parameter & Value & Source \\
 \hline
 \hline 
  $D_P$ & $2 \times 10^{-7}$ cm$^2$/s & \cite{martin1992applications,bera2022extracellular,young1980estimation} \\
  \hline
  $D_S$ & $2 \times 10^{-8}$ cm$^2$/s & \cite{martin1992applications,bera2022extracellular,masel1999quantifying,young1980estimation} \\
  \hline
  $a$ & $5 \times 10^{-5}$ s$^{-1}$ & \cite{masel1999quantifying} \\
  \hline
  $b$ & $9 \times 10^{-7}$ s$^{-1}$ & \cite{masel1999quantifying} \\
  \hline
  $S_c$ & $1 \times 10^{-4}$ g/cm$^3$ & \cite{beekes1996sequential,barber1970density} \\
  \hline
  $f$ & $0$ s$^{-1}$ & N/A \\
  \hline
  $c$ & $2$ $\frac{\text{cm}^3}{g s}$ & \cite{matthaus2006diffusion,barber1970density} \\
  \hline
  $t_d$ & $8 \times 10^{4}$ s & \cite{pena2011dengue} \\
  \hline
  $A$ & $2 \times 10^{-10}$ $\frac{\text{g}}{\text{cm}^2 \cdot \text{s}}^*$ & \cite{matthaus2006diffusion,barber1970density,braitenberg2001brain,raine} \\
  \hline  
  $R$ & $1 \times 10^{-2}$ cm & \cite{stuart2016dendrites} \\
  \hline  
\end{tabular}
\caption{Loose estimates of parameter values. $^*$: value reported is $4 \times 10^{5}$ times the true estimated flux of a single neuron in isolation --- see explanation in text.} \label{tab:param_vals}
\end{table}

\begin{table}
\centering
\begin{tabular}{c|c} 
 Parameter & Value \\
 \hline
 \hline 
  $\gamma$ & 4 \\
  \hline
  $\sigma$ & 0 \\
  \hline  
  $\delta$ & 0.1 \\
  \hline
  $\eta$ & 3.16 \\
  \hline 
  $\beta$ & $0.018$ \\
  \hline
  $\tau$ & 4 \\
  \hline
  ${\rho}$ & 0.158 \\
  \hline
\end{tabular}
\caption{Dimensionless values for clinical model.} \label{tab:ndim_vals}
\end{table}

\paragraph{Setup for Systems}

We focus on the behavior of our model with the estimated biological parameters. Numerical specifications for our biologically-driven study are given in Appendix \ref{app:bio}. The parameters of Table \ref{tab:param_vals} define characteristic scales of $\bar t = 20000$ s, $\bar x = 0.06$ cm, $\bar P = 6 \times 10^{-5}$ g/cm$^3$, and $\bar S = 0.0001$ g/cm$^3$. Throughout all simulations, we choose initial conditions so that the dimensional values of $i_c$ and $i_s$ of Eqs. \eqref{eq:ip} and \eqref{eq:is} are 
\begin{align} I_P &= \bar P \bar x^d \approx \begin{cases}
4 \times 10^{-6} \text{g/cm}^2, \quad d=1 \\
3 \times 10^{-7} \text{g/cm}, \quad d=2 \\
2 \times 10^{-8} \text{g}, \quad d=3
\end{cases} \label{eq:IPstar}  \\ 
I_S &= \bar S \bar x^d \approx \begin{cases}
6 \times 10^{-6} \text{g/cm}^2, \quad d=1 \\
4 \times 10^{-7} \text{g/cm}, \quad d=2 \\
3 \times 10^{-8} \text{g}, \quad d=3
\end{cases} \label{eq:ISstar} 
\end{align} when $i_p = i_s = 1.$ We also ensure the initial concentration profiles match exactly in dimensional units across all systems.

To retain a consistent dimensional point of comparison with identical initial conditions, we nondimensionalize according to the baselines parameters and simulate; this results in a slightly more general nondimensionalization discussed in Appendix \ref{app:ndim}.

\subsection{Sensitivity Analysis}

The capacity for medicine to modify some or most of the parameters seems likely. We therefore consider how small, 30\%, changes in each modifiable parameter in the system affect the UPR mechanism. Specifically, we look at $h_\text{avg}$ (Eq. \eqref{eq:damage}) and \begin{equation} W_\text{avg} := \omega_\text{avg}/ \bar t, \end{equation} the dimensional value of $\omega_\text{avg}$ (Eq. \eqref{eq:damage_om}). We comment that $h_\text{avg}$ is already dimensionless.

For baseline, we use the parameter values of Table \ref{tab:param_vals}. Then, we vary each parameter individually by 30\% up and down, keeping all others fixed at baseline. We assume cell sizes are fixed and do not change $R$. The results are found in Tables \ref{tab:sensitivity} and \ref{tab:W} in differing dimensions.

In the case $d=1$, the model has large $h_\text{avg}$-values, suggesting a very diseased state. In that case, even 30\% changes in many parameters do not result in large changes because so many parameters contribute to high S-presence. We notice much larger changes with $d=2$ where the S-presence is smaller. For $d=1$, the most beneficial changes were decreasing $A$, increasing $S_c$, increasing $b$, and increasing $D_S$. For $d=2$, the most beneficial changes were decreasing $A$, decreasing $c$, increasing $D_S$, and increasing $S_c.$ Both $d=1$ and $d=2$ show the same pattern of increase/decrease with the different parameter changes. The $d=3$ cases did not show notable disease prevalence with $h_\text{avg}=0$ and no oscillations.

As for the oscillation frequency, $W_\text{avg}$, most of the parameters had little effect. The only parameter that had an appreciable effect was the UPR delay $t_d.$ Oscillations tended not to occur much above $d=1$, but when present in $d=1$ and $d=2$, they tended to be very close in value.

\begin{table}
\centering
\begin{tabular}{c||c|c} 
 Parameter Changes & $h_\text{avg}$ (1D) & $h_\text{avg}$ (2D) \\
 \hline
 \hline 
 baseline & $0.830$ & $0.0687$  \\
 \hline 
 $D_P \gets 0.7 D_P$ & \color{red} $0.831$ (+0.1\%) & \color{red} $0.124$ (+80.5\%)   \\
 \hline 
 $D_P \gets 1.3 D_P$ & \color{blue} $0.828$ (-0.2\%) & \color{blue} $0.032$ (-53.7\%) \\
 \hline 
 $D_S \gets 0.7 D_S$ & \color{red} $0.842$ (+1.4\%) & \color{red} $0.187$ (+172\%) \\
 \hline 
 $D_S \gets 1.3 D_S$ & \color{blue} $0.812$ (-2.2\%)  & \color{blue} $0.0107$ (-84.4\%)  \\
 \hline
 $a \gets 0.7 a$ & \color{red} $0.840$ (+1.2\%)  & \color{red} $0.108$ (+57.2\%)   \\
 \hline
 $a \gets 1.3 a$ & \color{blue} $0.821$ (-1.1\%)  & \color{blue} $0.037$ (-46.1\%)   \\
 \hline 
 $b \gets 0.7 b$ & \color{red} $0.847$ (+2.0\%) & \color{red} $0.0947$ (+37.8\%)  \\
 \hline
 $b \gets 1.3 b$ & \color{blue} $0.810$ (-2.4\%)  & \color{blue} $0.047$ (-31.6\%)   \\
 \hline 
 $c \gets 0.7 c$ & \color{blue} $0.815$ (-1.8\%) & \color{blue} $0.00211$ (-96.9\%)  \\
 \hline 
 $c \gets 1.3 c$ &  \color{red} $0.838$ (+1.0\%) & \color{red} $0.140$ (+103.8\%)   \\
 \hline 
 $A \gets 0.7 A$ & \color{blue} $0.772$ (-7.0\%)  & \color{blue} $0.000128$ (-99.8\%)  \\
 \hline 
 $A \gets 1.3 A$ & \color{red} $0.860$ (+3.6\%)  & \color{red} $0.218$ (+217.3\%)  \\
 \hline
 $t_d \gets 0.7 t_d$ & \color{red} $0.836$ (+0.7\%) & \color{red} $0.0696$ (+1.3\%) \\
 \hline 
 $t_d \gets 1.3 t_d$ & \color{blue} $0.823$ (-0.8\%)  & \color{blue} $0.0678$  (-13.5\%) \\
 \hline
 $S_c \gets 0.7 S_c$ & \color{red} $0.861$ (+3.7\%) & \color{red} $0.160$ (+132.9\%)   \\
 \hline 
 $S_c \gets 1.3 S_c$ & \color{blue} $0.792$ (-4.6\%)  & \color{blue} $0.0197$ (-71.3\%)   \\
 \hline
\end{tabular} 
\caption{Values of $h_\text{avg}$ under various parameter changes with changes relative to baseline in parentheses. The $h_\text{avg}$ values were $0$ for all 3D cases.} \label{tab:sensitivity}
\end{table}

\begin{table}
\centering
\begin{tabular}{c||c|c} 
 Parameter Changes & $W_\text{avg}$ [$\mu$Hz] (1D) & $W_\text{avg}$ [$\mu$Hz] (2D) \\
 \hline
 \hline 
 baseline & $3.85$ & $0$  \\
 \hline 
 $D_P \gets 0.7 D_P$ & $3.91$ & $0$  \\
 \hline 
 $D_P \gets 1.3 D_P$ & $3.79$ & $0$ \\
 \hline 
 $D_S \gets 0.7 D_S$ & $3.79$ & $4.06$ \\
 \hline 
 $D_S \gets 1.3 D_S$ & $3.87$ & $0$   \\
 \hline
 $a \gets 0.7 a$ & $3.84$ & $0$  \\
 \hline
 $a \gets 1.3 a$ & $3.85$ & $0$  \\
 \hline 
 $b \gets 0.7 b$ & $3.78$ & $0$  \\
 \hline
 $b \gets 1.3 b$ & $3.80$ & $0$   \\
 \hline 
 $c \gets 0.7 c$ & $3.74$ & $0$  \\
 \hline 
 $c \gets 1.3 c$ & $3.92$ & $4.10$  \\
 \hline 
 $A \gets 0.7 A$ & $3.81$ & $0$  \\
 \hline 
 $A \gets 1.3 A$ & $3.87$ & $4.20$  \\
 \hline
 $t_d \gets 0.7 t_d$ & $5.19$ & $0$ \\
 \hline 
 $t_d \gets 1.3 t_d$ & $3.02$ & $0$  \\
 \hline
 $S_c \gets 0.7 S_c$ & $3.76$ & $4.03$  \\
 \hline 
 $S_c \gets 1.3 S_c$ & $3.88$ & $0$  \\
 \hline
\end{tabular} 
\caption{Values of $W_\text{avg}$ at various parameter changes. No oscillations were noted for any 3D cases.} \label{tab:W}
\end{table}

\subsection{Case Study}

It appears the drug pentosan polysulfate (PPS) can extend survival times in patients with prion diseases \cite{rainov2007experimental}. One man survived 10 years with CJD while receiving that drug treatment \cite{simms}. Researchers have suggested the drug inhibiting the binding of PrP$^\text{Sc}$ to PrP$^\text{C}$ or causing the fragmentation of PrP$^\text{Sc}$ as possible mechanisms \cite{yamasaki2014comparison}. It appears PPS does not affect levels of PrP$^\text{C}$ \cite{yamasaki2014comparison}. Other drugs have also been studied where PrP$^\text{C}$ levels were made significantly lower --- through chloropromazine (CPZ) and U18666A \cite{yamasaki2014comparison}.

\begin{figure}
\includegraphics[height=1.5in]{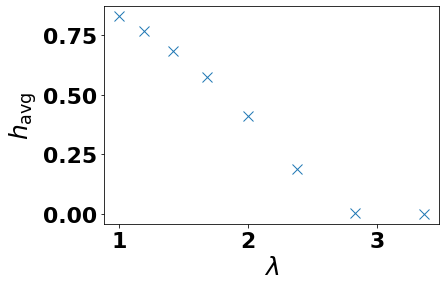} \includegraphics[height=1.5in]{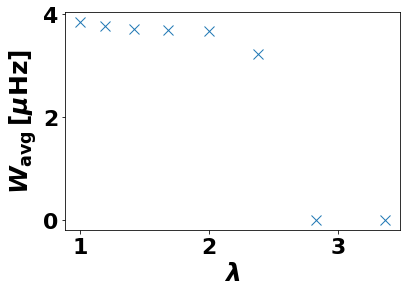}
\caption{Variations in $h_\text{avg}$ and $W_\text{avg}$ at various potencies.} \label{fig:drug_effects}
\end{figure}

To speculate quantitatively on drug treatments in our model, we consider a hypothetical concoction of drugs that (1) increases the clearance rate of PrP$^\text{Sc}$, (2) increases the diffusivity of PrP$^\text{Sc}$, (3) reduces the capacity of PrP$^\text{Sc}$ to convert PrP$^\text{C}$, and (4) reduces the production of PrP$^\text{C}$. Items (1)-(2) are inspired by the speculation on the PPS drug, where we assume fragmented PrP$^\text{Sc}$ will be cleared more readily and diffuse more easily. Item (3) is again a possible benefit of PPS, and item (4) is based on the possible effects of CPZ and U18666A. Numerically, we investigate how $h_\text{avg}$ and $W_\text{avg}$ change as the drug ``potency" changes. We define the potency $\lambda \in [1, \infty)$ so that at potency $\lambda$, the values of $D_s$ and $b$ are both increased by a factor $\lambda$ relative to baseline (Table \ref{tab:param_vals}) and the values of $c$ and $A$ are both decreased by a factor of $\lambda$ relative to baseline (Table \ref{tab:param_vals}). The resulting outcomes are found in Figure \ref{fig:drug_effects}. We observe that $h_\text{avg}$ is nearly zero and oscillations stop above $\lambda \approx 2.4.$

\subsection{Clinical Relevance}

At this point, we discuss the potential clinical significance of our model. From Table \ref{tab:sensitivity}, we note that the change in each dimensional parameter has a corresponding change in the strength of the UPR response $h_\text{avg}$. This suggests that medical interventions that increase the diffusivities of either or both proteins; reduce the P-S conversion rate; enhance the rate that either or both proteins are cleared; shorten the UPR delay; decrease the maximum flux of P; and increase the threshold at which the UPR is triggered, may benefit patients clinically. The most significant reductions in the UPR intensity seem to come about through decreasing the maximum rate P is released, decreasing the recruitment rate, increasing the diffusivity of S, increasing the rate at which S is cleared, and increasing the threshold sensitivity of the UPR to S. We note these results may vary depending on the point about which sensitivity is studied. We remark that based on the possible effects of drugs used in prion disease treatment to date, their methods of action are strongly aligned with these targets.

An experimentalist may be able to identify signs of the UPR mechanism through reduced production of PrP$^\text{C}$. Based on our simulations, we expect that oscillations will become more pronounced when the relative diffusivity of PrP$^\text{Sc}$ relative to PrP$^\text{C}$ is small, the relative clearance rate of PrP$^\text{Sc}$ relative to PrP$^\text{C}$ is small, and the UPR delay is large relative to the PrP$^\text{C}$ clearance time. It is also expected that the temporal frequency of oscillations will decrease as the UPR delay time increases and that for large enough diffusivities or clearances of PrP$^\text{Sc}$, the oscillations may entirely fade away.

\subsection{Model Limitations}

To arrive at the system we studied, we had to make several assumptions. A proper understanding of the resulting limitations is indispensable for healthy scientific scepticism in interpreting the model results and in developing better, more insightful models. We include a list of limitations below along with possible remedies.

\begin{itemize}
    \item {\bf Only two proteins:} instead of including multitudes of oligomeric states and stable nuclei, we chose to only include P- and S-proteins. A more refined model could consider assortments of assembly sizes, with associated mass-dependent diffusivities and clearance rates.
    \item {\bf Only one neuron:} we centered our model on one neuron, but it is possible to consider multiple neurons, with associated positions in space so that species that diffuse away from one neuron can affect another.
    \item {\bf Simplified neuron geometry:} we used simplified geometries for the neuron shape. Due to the long axons relative to the cell body size, radial symmetry is not accurate. Through a finite element approach, a more complicated geometry could be tackled.
    \item {\bf Constant reaction rates:} through the ageing process, the rates of misfolding or clearance, among other parameters, may vary. It is possible to consider time-dependent parameters as well.
    \item {\bf No membrane-bound P-protein:} we did not consider a form of P-protein that is bound to the cell membrane. Further refinements could add this. 
    \item {\bf Protein production not impacted by UPR:} the model could be adapted to include a reduction in maximal P-protein production due to cumulative effects of the UPR. Likewise, since the UPR may be a mechanism of cell death, the model could be extended to determine when the neuron dies.
    \item {\bf Only one cell type:} there are many cell types in the brain, but we focused only on neurons. By including other cell types, we could better model the effects of stress and inflammation.
    \item {\bf Lack of data-driven parameters:} due to the uncertainty of the different model parameters, we focused mostly on the qualitative aspects of the model. Our clinical exploration study provided interesting insights with regards to possible medical treatments, but accurate measurements of parameters are needed.
\end{itemize}

However, even with these limitations, our simple approach already highlights some important features of these complex mechanisms, as well as the role of key parameters involved.

\section{Conclusion and Future Work}

\label{sec:conclusion}

In this paper, we presented a nonlinear, coupled system of reaction-diffusion equations with nonlinear, delayed boundary conditions to model the Unfolded Protein Response in a simplified setting with representative healthy and unhealthy proteins. We found that oscillations in neuronal activity may be found under certain parameters and that through modifying certain biological parameters it may be possible to lessen intensity of the UPR.

To extend our work and make it more applicable to neurological systems and the study of neurodegenerative diseases, we consider obtaining accurate estimates for the model parameters based on clinical and experimental data, incorporating additional biological features and realism (see limitations section), and coming to a theoretical understanding of the effects of each model parameter.

\section*{Acknowledgments}

    This project has received support from NSF Award DMS \#2150478 and NSF Award DMS \#2316952.
This project has received support from Agence National de la Recherche PrionDiff 350 Project-ANR-21-CE15-0011. The authors would like to thank Erwin Suazo and Tamer Oraby for funding students in this work.

\appendix

\section{Numerical Method}

\label{app:num}

\subsection{Different Nondimensionalization}

\label{app:ndim}

With scales $\bar t,$ $\bar x,$ $\bar P,$ and $\bar S$ chosen from the baseline parameters, the dimensionless system, more generally, takes the form

\begin{align}
    p_{,t} &= \delta_p (p_{,rr} + \frac{d-1}{r} p_{,r} ) - \gamma ps - (\alpha+\sigma) p, \\
    s_{,t} &= \delta_s ( s_{,rr} + \frac{d-1}{r} s_{,r} ) + \eta \gamma ps + \eta \sigma p - \beta s,  \\
    p_{,r}|_{r=\rho} &= \frac{-\theta}{1+ (\mu s(\rho,t-\tau))^m} \\
    h_\text{avg} &= \lim_{\Upsilon \to \infty} \int_0^\Upsilon (1 - \frac{1}{1 + (\mu s(\rho, t-\tau))^m} ) \dd t,  
\end{align}
where we have suppressed equations that do not change their form. Here, $\delta_p = \frac{\bar t D_P}{\bar x^2},$ $\gamma = \bar S \bar t c,$ $\alpha = \bar t a,$ $\sigma = \bar t f,$ $\eta = \frac{\bar P}{\bar S},$ $\delta_s = \frac{\bar t D_S}{\bar x^2},$ $\beta = \bar t b,$ $\rho= \frac{R}{\bar x},$ $\theta = \frac{\bar x A}{\bar P D_P},$ $\mu = \frac{\bar S}{S_c}$, and  $\tau = \frac{t_d}{\bar t}.$ With the baseline parameters, $\delta_p = \alpha = \theta = \mu = 1.$ 

\subsection{Implementation}

\label{app:imp}

We solve the dimensionless system of Appendix \ref{app:ndim}. The domain for $r$, $[\rho,\infty),$ is unbounded and we need to truncate the computational domain at some value, $r_\infty.$ From a back-of-the-envelope calculation, we note that for $s=0$, a one-dimensional steady state solution is $p(r) = \frac{1}{\sqrt{1+\sigma}} \exp(-\sqrt{1+\sigma}(r-\rho)) \leq \exp(-(r-\rho)).$ And if we impose $p(r_\infty)=0$ with $r_\infty=\rho+7$, the truncation error is $\sim 0.0009.$ And $p$ and $s$ are coupled together, with $p$ being the only source for $s$. Solutions tend to decay faster in $2$ and $3$ dimensions. Thus, for $O(1)$ solutions and data, choosing $r_\infty=\rho+7$, we anticipate associated truncation errors to be $\sim 0.001$ in the far field.

We choose positive integers $N_r$ and $N_t$ to be the meshing parameters in space and time. For a computational domain running from $r=\rho$ to $r=r_\infty$ and $t=0$ to $t=t_\infty,$ we define the spatial and temporal step sizes by
\begin{align*}
    &\Delta r = \frac{r_\infty - \rho}{N_r},& & \Delta t = \frac{t_\infty}{N_t}.&
\end{align*}
Then for $i=0,1,2,...,N_r$ and $j=0,1,...,N_t$, we define $r_i = \rho+i \Delta r$ and $t_j = j \Delta t$. We denote $p_i^j$ as the numerical approximation to $p(x_i,t_j)$ and $s_i^j$ as the numerical approximation to $s(r_i,t_j).$ We furthermore take the delay $\tau=k \Delta t$ for some $k \in \{0,1,2, \ldots\}.$ At step $j$, second-order spatial discretizations of the diffusion and reaction terms for $p$ and $s$ are given by 
\begin{align}
    \bar M_p &= \delta_p \frac{p_{i+1}^{j+1}-2p_i^{j+1}+p_{i-1}^{j+1}}{\Delta r^2} - \gamma p_i^{j+1}s_i^j \nonumber \\ &- (1+\sigma)p_i^{j+1}, 0 \leq i \leq N_r - 1, \quad j = 0, 1, 2, ..., \\
    \bar M_s &= \delta_s \frac{s_{i+1}^{j+1}-2s_i^{j+1}+s_{i-1}^{j+1}}{\Delta r^2} + \eta \gamma p_i^js_i^{j+1} \nonumber \\ &+ \eta \sigma p_i^{j} - \beta s_i^{j+1}, 0 \leq i \leq N_r - 1, \quad j = 0, 1, 2, ...,
\end{align}
respectively.

Handling the derivative conditions at $i=0$ requires ghost points at the fictitious position $r_{-1} = \rho - \Delta r$, and the far-field at $i=N_r$, zero values are imposed. Thus:
\begin{align}
        p_{N_r}^j &= s_{N_r}^j = 0, \quad j = 0, 1, 2, ... \\
    \frac{p_{1}^{j}-p_{-1}^{j}}{2\Delta r} &= \frac{-\theta}{1 + {(\mu s_0^{j-k})}^m}, \quad j = 0, 1, 2, ... \\
    \frac{s_{1}^{j}-s_{-1}^{j}}{2\Delta r} &= 0, \quad j=0, 1, 2, ...\,.
\end{align}

To handle the delay, we assume $s_0^{-1}, s_0^{-2}, \ldots, s_0^{-k}$ are given where $\tau = k \Delta t.$ For initial conditions, we assume ${p_i}^0, {s_i}^0$ are given for $0 \leq i \leq N_r$. 

These equations can be converted to matrix-vector form. We denote $p^j = (p_0^j, p_1^j, ..., p_{N_r-1}^j)^\intercal$ and $s^j = (s_0^j, s_1^j, ..., s_{N_r-1}^j)^\intercal$. Let $I \in \mathbb{R}^{N_r \times N_r}$ be the identity matrix, $D_2 \in \mathbb{R}^{N_r \times N_r}$ be given by $$D_2 = \frac{1}{\Delta r ^ 2}\begin{pmatrix}
 -2 & 2 & 0 & \cdots & 0 & 0 \\
1 & -2 & 1 & \cdots & 0 & 0 \\
0 & \ddots & \ddots & \ddots & \vdots & \vdots \\
\vdots & \cdots & \ddots & \ddots & \ddots & \vdots \\
0 & 0 & \cdots & 1 & -2 & 1 \\
0 & 0 & \cdots & 0 & 1 & -2
\end{pmatrix},
$$
$D_1 \in \mathbb{R}^{N_r \times N_r}$ be given by $$D_1 = \begin{cases}    
 \frac{(d - 1)}{2 \Delta r}\begin{pmatrix}
 0 & 0 & 0 & \cdots & 0 & 0 \\
-1/r_1 & 0 & 1/r_1 & \cdots & 0 & 0 \\
0 & \ddots & \ddots & \ddots & \vdots & \vdots \\
\vdots & \cdots & \ddots & \ddots & \ddots & \vdots \\
0 & 0 & \cdots & -1/r_{N_r-2} & 0 & 1/r_{N_r-2} \\
0 & 0 & \cdots & 0 & -1/r_{N_r-1} & 0
\end{pmatrix}
, \quad d = 2, 3, \\
0, \quad d=1,
\end{cases}$$ and $V^j \in \mathbb{R}^{N_r \times 1}$ be given by
$$V^j = (\frac{2 \theta \delta_p}{\Delta r (1 + (\mu {s_0}^{j-k})^m)}, 0, 0, \ldots, 0)^\intercal + \begin{cases}(\frac{-\theta \delta_p (d-1)}{\rho(1 + (\mu {s_0}^{j-k})^m)}, 0, 0, \ldots, 0)^\intercal, \quad d=2,3, \\ 
0, \quad d=1. \end{cases}$$ Defining functions $M_p, M_s : \mathbb{R}^{N_r \times 1} \to \mathbb{R}^{N_r \times N_r}$ via 
\begin{align*}
M_p(s) &= \delta_p( D_2 + D_1) - \gamma \diag(s) - (\sigma+\alpha)I \\
M_s(p) &= \delta_s (D_2 + D_1) + \eta \gamma \diag(p) - \beta I,
\end{align*}
we can define a weighted semi-implicit first-order time-stepping via
\begin{align*}
\frac{p^{j+1}-p^j}{\Delta t} &= w (M_p(s^j) p^{j+1} + V^{j+1}) + (1-w) (M_p(s^j) p^j + V^j) \\
\frac{s^{j+1}-s^j}{\Delta t} &= w M_s(p^j) s^{j+1} + (1-w) M_s(p^j) s^j + \eta \sigma p^j
\end{align*}
where $0 \leq w \leq 1.$ With $w=0$ this is the Euler method. With $w=1$, this is a standard semi-implicit method. We take inspiration from the implicit Crank-Nicholson method \cite{douglas1961survey} and take $w=1/2.$ In practice, we find $w=1/2$ yields smaller errors with our semi-implicit approach. The semi-implicit systems require simple linear solves of two systems of size $N_r$. In contrast, Crank-Nicholson would require nonlinear solves of systems of size $2N_r$ and we find that the semi-implicit method is faster. When $k=0$ (no delay), we compute $V^{j+1}$ by first estimating $s^{j+1}$ with the Improved Euler method.

Overall our method, as subsequently validated in Appendix \ref{app:valid}, is second order in space and first order in time.

\subsection{Frequency}

\label{app:freq}

To estimate the mean frequency over a finite computational window, we make some modifications. Because $s$ can take a while to peak and a final peak may take place sufficiently before $t_\infty$ to contribute a non-negligible error, we seek to estimate the mean peak-to-peak frequency of $s(\rho,t-\tau).$ But because oscillations may stop altogether, we need to avoid giving a frequency for situations where $s(\rho,t-\tau)$ peaks a few times early on and then stops. If there $N \leq 1$ peaks for $s(\rho,t-\tau)$ on $(0,t_\infty)$, we define $\omega_\text{avg}=0$ since a mean peak-to-peak interval cannot be computed for the frequency. Assuming there are $N \geq 2$ peaks, let them occur at the  times $0 < t_1 < t_2 < ... < t_N < t_\infty.$ We define a proxy for the mean frequency as $$\hat \omega = \frac{N-1}{t_N-t_1}.$$ Then we define
\begin{equation}
\omega_\text{avg} = \begin{cases} \hat \omega, \quad t_\infty \leq t_N + \max_{1 \leq i \leq N-1} (t_{i+1}-t_i)\\
0, \quad \text{ otherwise.}
\end{cases}
\end{equation}
Intuitively, if the system is continuing to oscillate, then the end of the computational time $t_\infty,$ should occur before the estimated latest possible time for the $(N+1)$st peak (otherwise there should have been $N+1$ peaks). If this condition fails, we define the frequency to be $0$, indicating the system stopped oscillating.

\subsection{Validation of Numerical Scheme}

\label{app:valid}

Throughout the validation and simulations, the initial conditions prescribed are those of Equations \eqref{eq:pic}-\eqref{eq:pastic}.

To validate the convergence rates, we shall use mesh refinements. Let $N_r$ and $N_t$ be the meshing parameters. Denote $Q(N_r,N_t)_i^j = (p_i^j, q_i^j)^\intercal$ for $0 \leq i \leq N_r-1, 0 \leq j \leq N_t \}$ to be the numerical approximation at $r=r_i$ and $t=t_j$ over the chosen grid. We define the operations
\begin{align*}
\Lambda_r Q(N_r, N_t) &= \max_{\substack{0 \leq i \leq N_r-1 \\ 0 \leq j \leq N_t}} ||Q(N_r,N_t)_i^j - Q(2N_r,N_t)_{2i}^j||_\infty \\
\Lambda_t Q(N_r, N_t) &= \max_{\substack{0 \leq i \leq N_r-1 \\ 0 \leq j \leq N_t}} ||Q(N_r,N_t)_i^j - Q(N_r,2N_t)_{i}^{2j}||_\infty 
\end{align*}
If $q_i^j = q(r_i, t_j) = (p(r_i,t_j), s(r_i,t_j))^\intercal$ is the exact solution on the same mesh as $Q(N_r,N_t)$ then in the asymptotic limit, we anticipate that $$\max_{i,j} ||Q(N_r,N_t)_i^j - q_i^j||_\infty = \mathcal{O}(\Delta r^2) + \mathcal{O}(\Delta t)$$ and we make the ansatz that $$Q(N_r,N_t)_i^j - q_i^j = H(r_i,t_j)/N_r^2 + K(r_i,t_j)/N_t$$ for $O(1)$ error functions $H$ and $K$. From this, we have \begin{align*} Q(N_r,N_t)_i^j - Q(2N_r,N_t)_{2i}^j &= \frac{3H(r_i,t_j)}{4N_r^2} \\ Q(N_r,N_t)_i^{j} - Q(N_r,2N_t)_i^{2j} &= \frac{K(r_i,t_j)}{2 N_t}.\end{align*} Finally, assuming the meshes are refined enough, we have that 
\begin{align*}
\Lambda_r Q(N_r, N_t) &= \frac{3H^*}{4 N_r^2} \\
\Lambda_t Q(N_r, N_t) &= \frac{K^*}{2 N_t} 
\end{align*}
where $H^* = \sup_{[\rho,r_\infty]\times[0,t_\infty]} ||H||_\infty$ and $K^* = \sup_{[\rho,r_\infty]\times[0,t_\infty]} ||K||_\infty.$

In our validation, we focus our attention on simulations near the biologically relevant parameters of Table \ref{tab:ndim_vals} and the phase diagrams of Figures \ref{fig:phase0}-\ref{fig:phase4}. With $t_\infty = 80$, we choose $\gamma=5$, $\sigma=0.01$, $\delta = 0.05$, $\beta=0.01$, $\eta=1$, $\tau=5$, and $\rho=0.15.$

\begin{figure}
\centering
\includegraphics[width=0.45\linewidth]{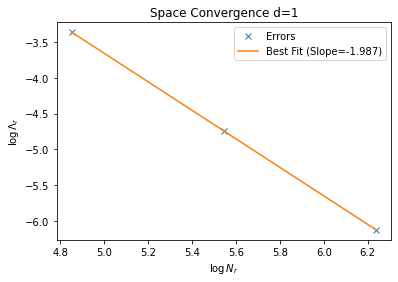} \includegraphics[width=0.45\linewidth]{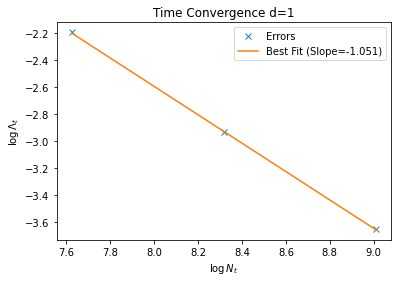} \\
\includegraphics[width=0.45\linewidth]{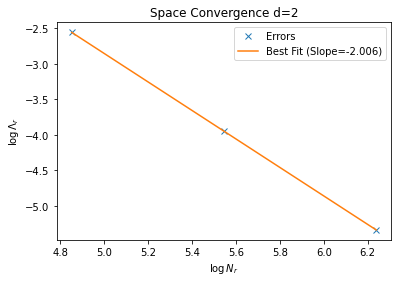} \includegraphics[width=0.45\linewidth]{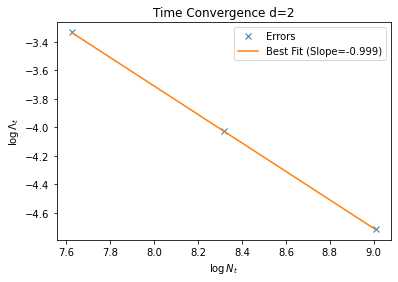} \\
\includegraphics[width=0.45\linewidth]{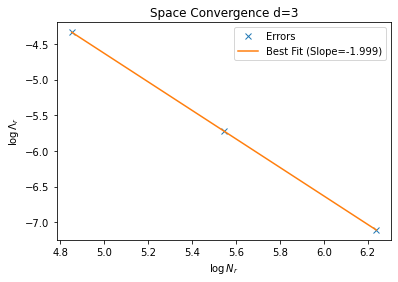} \includegraphics[width=0.45\linewidth]{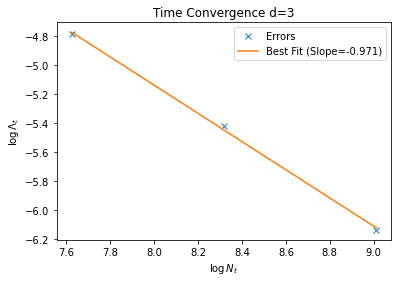} 
\caption{Plots of convergence with best fit.} \label{fig:conv}
\end{figure}

Numerically, we fix $N_t = 2^{14}$ and vary $N_r$ over the values $2^i$ where $i=7,8,9,10$ to compute $\Lambda_r$-values. Then we fix $N_r = 2^{10}$ and vary $N_t$ over the values $2^j$ where $j=11,12,13,14$ to compute $\Lambda_t$-values. We have
\begin{align*}
\log \Lambda_r Q(N_r, N_t) &= ( \log(3/4) + \log(H^*) ) - 2 \log N_r &&= b_r +  m_r \log N_r \\
\log \Lambda_t Q(N_r, N_t) &= ( \log(1/2) + \log(K^*) ) - \log N_t &&= b_t + m_t \log N_t
\end{align*}
for slopes $m_r$ and $m_t$ we expect to be close to $-2$ and $-1$, respectively, and intercepts $b_r$ and $b_t.$ The values $H^*$ and $K^*$ can be found from $b_r$ and $b_t$. From lines of best fit, we estimate these values in Tables \ref{tab:space_conv} and \ref{tab:time_conv}. Convergence plots are found in Figure \ref{fig:conv}.

\begin{table}
\centering
\begin{tabular}{c|c|c|c} 
$d$ & $m_r$ & $b_r$ & $H^*$ \\
 \hline
 \hline 
$1$ & $-1.987$ & $6.275$ & $708.292$ \\
\hline
$2$ & $-2.006$ & $7.177$ & $1744.776$ \\
\hline
$3$ & $-1.999$ & $5.362$ & $284.106$ 
\end{tabular} 
\caption{Spatial convergence results.} \label{tab:space_conv}
\end{table}

\begin{table}
\centering
\begin{tabular}{c|c|c|c} 
$d$ & $m_t$ & $b_t$ & $K^*$\\
 \hline
 \hline 
$1$ & $-1.051$ & $5.812$ & $668.522$ \\
\hline
$2$ & $-0.999$ & $4.282$ & $144.800$ \\
\hline
$3$ & $-0.971$ & $2.629$ & $27.720$ 
\end{tabular} 
\caption{Temporal convergence results} \label{tab:time_conv}
\end{table}

These results suggest that to keep the asymptotic errors in both space and time below $0.005$, we can use meshes as presented in Table \ref{tab:meshes}.

\begin{table}
\centering
\begin{tabular}{c|c|c} 
$d$ & $N_t$ & $N_r$ \\
 \hline
 \hline 
$1$ & $150000$ & $400$ \\
\hline
$2$ & $30000$ & $600$ \\
\hline
$3$ & $6000$ & $300$ 
\end{tabular} 
\caption{Estimates of mesh parameters to keep asymptotic errors $\leq 0.005$.} \label{tab:meshes}
\end{table}

\section{Numerical Specifications}

\label{app:spec}

\subsection{Model Investigation}

\label{app:model}

\subsubsection{Phase Diagram (Figures \ref{fig:phase0}-\ref{fig:phase4})}

For a computational domain, we used $[\rho, r_\infty]$ with $\rho=0.25$ and $r_\infty=7.25$ for $r$, and $[0,t_\infty]$ with $t_\infty=80$ for $t$. We used $r_* = (3\rho+r_\infty)/4$, $i_p = i_s = n_p = 1.$ Our meshing parameters were those presented in Table \ref{tab:meshes}. The switch parameter $m=10$ was used.

\subsubsection{Spatiotemporal Parameter Variations (Figures \ref{fig:pbehavior}-\ref{fig:extra})}

For a computational domain, we used $[\rho, r_\infty]$ (with $\rho$ as specified in the figures) with $r_\infty=\rho+7$ for $r$, and $[0,t_\infty]$ with $t_\infty=80$ for $t$. We used $r_* = (3\rho+r_\infty)/4$, $i_p = i_s = n_p = 1.$ To ensure similar accuracy as in the phase diagram, we used meshing parameters outlined in Table \ref{tab:meshes_vary}. The switch parameter $m=10$ was used.

\begin{table}
\centering
\begin{tabular}{c|c|c} 
$d$ & $N_t$ & $N_r$ \\
 \hline
 \hline 
$1$ & $140000$ & $400$ \\
\hline
$2$ & $120000$ & $500$ \\
\hline
$3$ & $80000$ & $300$ 
\end{tabular} 
\caption{Estimates of mesh parameters to keep asymptotic errors $\leq 0.005$.} \label{tab:meshes_vary}
\end{table}

\subsubsection{Comparison with Prior Work (Figure \ref{fig:compare})}

For a computational domain, we used $[0.25, r_\infty]$ with $r_\infty=\rho+7$ for $r$, and $[0,t_\infty]$ with $t_\infty=80$ for $t$. We used $r_* = (3\rho+r_\infty)/4$, $i_p = i_s = n_p = 1.$ We used meshing parameters outlined in Table \ref{tab:meshes}. The switch parameter $m=10$ was used.

\subsection{Biological Investigation}

\label{app:bio}

\subsubsection{Generation of Tables \ref{tab:sensitivity}-\ref{tab:W} and Figure \ref{fig:drug_effects}}

For a computational domain, we used $[\rho, r_\infty]$ with $\rho$ found in Table \ref{tab:ndim_vals} and $r_\infty=\rho+7$ for $r$ and $[0,t_\infty]$ with $t_\infty=80$ for $t$. We used $r_* = (3\rho+r_\infty)/4.$ Our meshing parameters were those presented in Table \ref{tab:meshes}. The switch parameter $m=10$ was used. In this application, $i_p = i_s = 1$ again, but $n_p$ varied depending on the parameter that changed.

\bibliographystyle{vancouver}

\begin{thebibliography}{10}

\bibitem{goedert2006century}
Goedert M, Spillantini MG.
\newblock {A century of Alzheimer's disease}.
\newblock science. 2006;314(5800):777-81.

\bibitem{davie2008review}
Davie CA.
\newblock {A review of Parkinson's disease}.
\newblock British medical bulletin. 2008;86(1):109-27.

\bibitem{akhtar2021neurodegenerative}
Akhtar A, Andleeb A, Waris TS, Bazzar M, Moradi AR, Awan NR, et~al.
\newblock {Neurodegenerative diseases and effective drug delivery: A review of
  challenges and novel therapeutics}.
\newblock Journal of Controlled Release. 2021;330:1152-67.

\bibitem{kiaei2013new}
Kiaei M.
\newblock New hopes and challenges for treatment of neurodegenerative
  disorders: Great opportunities for young neuroscientists.
\newblock Basic and Clinical Neuroscience. 2013;4(1):3.

\bibitem{newby2013blessings}
Newby GA, Lindquist S.
\newblock Blessings in disguise: biological benefits of prion-like mechanisms.
\newblock Trends in cell biology. 2013;23(6):251-9.

\bibitem{ono2009structure}
Ono K, Condron MM, Teplow DB.
\newblock Structure--neurotoxicity relationships of amyloid $\beta$-protein
  oligomers.
\newblock Proceedings of the National Academy of Sciences.
  2009;106(35):14745-50.

\bibitem{tolar2020path}
Tolar M, Abushakra S, Sabbagh M.
\newblock {The path forward in Alzheimer's disease therapeutics: Reevaluating
  the amyloid cascade hypothesis}.
\newblock Alzheimer's \& Dementia. 2020;16(11):1553-60.

\bibitem{stefanis2012alpha}
Stefanis L.
\newblock {$\alpha$-Synuclein in Parkinson's disease}.
\newblock Cold Spring Harbor perspectives in medicine. 2012;2(2):a009399.

\bibitem{prusiner1989creutzfeldt}
Prusiner S.
\newblock {Creutzfeldt-Jakob disease and scrapie prions}.
\newblock Alzheimer Disease \& Associated Disorders. 1989;3(1):52-78.

\bibitem{hetz2020mechanisms}
Hetz C, Zhang K, Kaufman RJ.
\newblock Mechanisms, regulation and functions of the unfolded protein
  response.
\newblock Nature reviews Molecular cell biology. 2020;21(8):421-38.

\bibitem{halliday2014targeting}
Halliday M, Mallucci GR.
\newblock Targeting the unfolded protein response in neurodegeneration: a new
  approach to therapy.
\newblock Neuropharmacology. 2014;76:169-74.

\bibitem{fricker2018neuronal}
Fricker M, Tolkovsky AM, Borutaite V, Coleman M, Brown GC.
\newblock Neuronal cell death.
\newblock Physiological reviews. 2018;98(2):813-80.

\bibitem{apodaca2006cellular}
Apodaca J, Kim I, Rao H.
\newblock Cellular tolerance of prion protein PrP in yeast involves proteolysis
  and the unfolded protein response.
\newblock Biochemical and biophysical research communications.
  2006;347(1):319-26.

\bibitem{hao2016mathematical}
Hao W, Friedman A.
\newblock {Mathematical model on Alzheimer’s disease}.
\newblock BMC systems biology. 2016;10(1):1-18.

\bibitem{puri2010mathematical}
Puri IK, Li L.
\newblock {Mathematical modeling for the pathogenesis of Alzheimer's disease}.
\newblock PloS one. 2010;5(12):e15176.

\bibitem{lindstrom2021reaction}
Lindstrom MR, Chavez MB, Gross-Sable EA, Hayden EY, Teplow DB.
\newblock {From reaction kinetics to dementia: A simple dimer model of
  Alzheimer’s disease etiology}.
\newblock PLoS computational biology. 2021;17(7):e1009114.

\bibitem{desplats2009inclusion}
Desplats P, Lee HJ, Bae EJ, Patrick C, Rockenstein E, Crews L, et~al.
\newblock Inclusion formation and neuronal cell death through neuron-to-neuron
  transmission of $\alpha$-synuclein.
\newblock Proceedings of the National Academy of Sciences.
  2009;106(31):13010-5.

\bibitem{bakshi2019mathematical}
Bakshi S, Chelliah V, Chen C, van~der Graaf PH.
\newblock {Mathematical biology models of Parkinson's disease}.
\newblock CPT: pharmacometrics \& systems pharmacology. 2019;8(2):77-86.

\bibitem{pandya2019predictive}
Pandya S, Zeighami Y, Freeze B, Dadar M, Collins DL, Dagher A, et~al.
\newblock Predictive model of spread of Parkinson's pathology using network
  diffusion.
\newblock NeuroImage. 2019;192:178-94.

\bibitem{greer2006mathematical}
Greer ML, Pujo-Menjouet L, Webb GF.
\newblock A mathematical analysis of the dynamics of prion proliferation.
\newblock Journal of theoretical biology. 2006;242(3):598-606.

\bibitem{salman2018mathematical}
Salman S, Ahmed E.
\newblock A mathematical model for Creutzfeldt Jacob Disease (CJD).
\newblock Chaos, Solitons \& Fractals. 2018;116:249-60.

\bibitem{adimy2022neuron}
Adimy M, Babin L, Pujo-Menjouet L.
\newblock {Neuron Scale Modeling of Prion Production with the Unfolded Protein
  Response}.
\newblock SIAM Journal on Applied Dynamical Systems. 2022;21(4):2487-517.

\bibitem{adimy2023multigroup}
Adimy M, Chekroun A, Pujo-Menjouet L, Sensi M.
\newblock A multigroup approach to delayed prion production.
\newblock Discrete and Continuous Dynamical Systems - Series B. 2024;29.

\bibitem{trusina2010unfolded}
Trusina A, Tang C.
\newblock The unfolded protein response and translation attenuation: a
  modelling approach.
\newblock Diabetes, Obesity and Metabolism. 2010;12:27-31.

\bibitem{goriely2020neuronal}
Goriely A, Kuhl E, Bick C.
\newblock Neuronal oscillations on evolving networks: dynamics, damage,
  degradation, decline, dementia, and death.
\newblock Physical review letters. 2020;125(12):128102.

\bibitem{garzon2021dynamics}
Garz{\'o}n DN, Castillo Y, Navas-Zuloaga MG, Darwin L, Hardin A, Culik N,
  et~al.
\newblock Dynamics of prion proliferation under combined treatment of
  pharmacological chaperones and interferons.
\newblock Journal of Theoretical Biology. 2021;527:110797.

\bibitem{atkinson2016prion}
Atkinson CJ, Zhang K, Munn AL, Wiegmans A, Wei MQ.
\newblock Prion protein scrapie and the normal cellular prion protein.
\newblock Prion. 2016;10(1):63-82.

\bibitem{selkoe2003folding}
Selkoe DJ.
\newblock Folding proteins in fatal ways.
\newblock nature. 2003;426(6968):900-4.

\bibitem{shafiq2022prion}
Shafiq M, Da~Vela S, Amin L, Younas N, Harris DA, Zerr I, et~al.
\newblock {The prion protein and its ligands: Insights into structure-function
  relationships}.
\newblock Biochimica et Biophysica Acta (BBA)-Molecular Cell Research.
  2022;1869(6):119240.

\bibitem{rangel2013non}
Rangel A, Race B, Striebel J, Chesebro B.
\newblock Non-amyloid and amyloid prion protein deposits in prion-infected mice
  differ in blockage of interstitial brain fluid.
\newblock Neuropathology and applied neurobiology. 2013;39(3):217-30.

\bibitem{moreno2012sustained}
Moreno JA, Radford H, Peretti D, Steinert JR, Verity N, Martin MG, et~al.
\newblock Sustained translational repression by eIF2$\alpha$-P mediates prion
  neurodegeneration.
\newblock Nature. 2012;485(7399):507-11.

\bibitem{di2022disturbance}
Di~Domenico F, Lanzillotta C.
\newblock Chapter Two - The disturbance of protein synthesis/degradation
  homeostasis is a common trait of age-related neurodegenerative disorders.
\newblock Advances in Protein Chemistry and Structural Biology. 2022;132:49-87.

\bibitem{martin1992applications}
Martin JB.
\newblock Applications of molecular genetics to restorative neurology.
\newblock In: Principles and Practice of Restorative Neurology. Elsevier; 1992.
  p. 189-201.

\bibitem{bera2022extracellular}
Bera K, Kiepas A, Godet I, Li Y, Mehta P, Ifemembi B, et~al.
\newblock Extracellular fluid viscosity enhances cell migration and cancer
  dissemination.
\newblock Nature. 2022;611(7935):365-73.

\bibitem{young1980estimation}
Young M, Carroad P, Bell R.
\newblock Estimation of diffusion coefficients of proteins.
\newblock Biotechnology and bioengineering. 1980;22(5):947-55.

\bibitem{hrabe2004model}
Hrabe J, Hrab{\u{e}}tov{\'a} S, Segeth K.
\newblock A model of effective diffusion and tortuosity in the extracellular
  space of the brain.
\newblock Biophysical journal. 2004;87(3):1606-17.

\bibitem{masel1999quantifying}
Masel J, Jansen VA, Nowak MA.
\newblock Quantifying the kinetic parameters of prion replication.
\newblock Biophysical chemistry. 1999;77(2-3):139-52.

\bibitem{beekes1996sequential}
Beekes M, Baldauf E, Diringer H.
\newblock Sequential appearance and accumulation of pathognomonic markers in
  the central nervous system of hamsters orally infected with scrapie.
\newblock Journal of General Virology. 1996;77(8):1925-34.

\bibitem{barber1970density}
Barber TW, Brockway JA, Higgins LS.
\newblock The density of tissues in and about the head.
\newblock Acta neurologica scandinavica. 1970;46(1):85-92.

\bibitem{matthaus2006diffusion}
Matth{\"a}us F.
\newblock Diffusion versus network models as descriptions for the spread of
  prion diseases in the brain.
\newblock Journal of theoretical biology. 2006;240(1):104-13.

\bibitem{pena2011dengue}
Pe{\~n}a J, Harris E.
\newblock Dengue virus modulates the unfolded protein response in a
  time-dependent manner.
\newblock Journal of Biological Chemistry. 2011;286(16):14226-36.

\bibitem{braitenberg2001brain}
Braitenberg V.
\newblock Brain size and number of neurons: an exercise in synthetic
  neuroanatomy.
\newblock Journal of computational neuroscience. 2001;10:71-7.

\bibitem{raine}
Raine C. Characteristics of the Neuron; 1999.
\newblock https://www.ncbi.nlm.nih.gov/books/NBK28209/.

\bibitem{stuart2016dendrites}
Stuart G, Spruston N, H{\"a}usser M.
\newblock Dendrites.
\newblock Oxford University Press; 2016.

\bibitem{rainov2007experimental}
Rainov N, Tsuboi Y, Krolak-Salmon P, Vighetto A, Doh-Ura K.
\newblock Experimental treatments for human transmissible spongiform
  encephalopathies: is there a role for pentosan polysulfate?
\newblock Expert opinion on biological therapy. 2007;7(5):713-26.

\bibitem{simms}
Wade J.
\newblock Belfast man suffering from variant CJD passes away.
\newblock The Journal. 2011.
\newblock Available from:
  \url{https://www.thejournal.ie/belfast-man-suffering-from-variant-cjd-passes-away-100653-Mar2011/}.

\bibitem{yamasaki2014comparison}
Yamasaki T, Suzuki A, Hasebe R, Horiuchi M.
\newblock Comparison of the anti-prion mechanism of four different anti-prion
  compounds, anti-PrP monoclonal antibody 44B1, pentosan polysulfate,
  chlorpromazine, and U18666A, in prion-infected mouse neuroblastoma cells.
\newblock PLoS One. 2014;9(9):e106516.

\bibitem{douglas1961survey}
Douglas~Jr J.
\newblock A survey of numerical methods for parabolic differential equations.
\newblock In: Advances in computers. vol.~2. Elsevier; 1961. p. 1-54.

\end{thebibliography}

\end{document}